\documentclass[
reprint,
showkeys,
longbibliography,
nofootinbib,
superscriptaddress,
aps,
notitlepage
]{revtex4-1}
\usepackage[utf8]{inputenc}

\usepackage{amsmath}
\usepackage{amssymb}
\usepackage{bm}
\usepackage{enumitem}
\usepackage[acronym]{glossaries}
\usepackage{graphicx}
\usepackage[colorlinks,unicode]{hyperref}
\usepackage[capitalise]{cleveref}  % must come after hyperref
\usepackage{mathtools}
\usepackage{multirow}
\usepackage{physics}
\usepackage{siunitx}
\usepackage[dvipsnames]{xcolor}

\DeclareSIUnit\week{week}

\makeglossaries
\newacronym{dm}{DM}{dark matter}
\newacronym{ce}{CE}{Cosmic Explorer}
\newacronym{et}{ET}{Einstein Telescope}
\newacronym{snr}{SNR}{signal-to-noise ratio}

\makeatletter
\renewcommand\onecolumngrid{%
\do@columngrid{one}{\@ne}%
\def\set@footnotewidth{\onecolumngrid}%
\def\footnoterule{\kern-6pt\hrule width 1.5in\kern6pt}%
}
\makeatother

\DeclareSymbolFont{mathtx}{OML}{txmi}{m}{it}
\DeclareMathAlphabet\mathbfcal{OMS}{cmsy}{b}{n}
\DeclareMathSymbol{v}{\mathalpha}{mathtx}{118}

\newcommand\myshade{80}
\colorlet{mylinkcolor}{ForestGreen}
\colorlet{mycitecolor}{Red}
\colorlet{myurlcolor}{violet}

\hypersetup{
  linkcolor  = mylinkcolor!\myshade!black,
  citecolor  = mycitecolor!\myshade!black,
  urlcolor   = myurlcolor!\myshade!black,
  colorlinks = true
}

\newcommand{\NbodyIMRI}{\texttt{NbodyIMRI}}
\newcommand{\risco}{R_\mathrm{ISCO}}
\newcommand{\vrel}{V_0}
\newcommand{\vrelvec}{\bm{V}_0}
\newcommand{\vrelhatvec}{\hat{\bm{V}}_0}

\DeclareSIUnit\solarmass{\ensuremath{\mathrm{M}_\odot}}
\DeclareSIUnit\parsec{pc}
\DeclareSIUnit\year{yr}

\newcommand{\GRAPPA}{Gravitation Astroparticle Physics Amsterdam (GRAPPA),\\ Institute for Theoretical Physics Amsterdam and Delta Institute for Theoretical Physics,\\ University of Amsterdam, Science Park 904, 1098 XH Amsterdam, The Netherlands}

\newcommand{\IFCA}{Instituto de F\'isica de Cantabria (IFCA, UC-CSIC), Av.~de Los Castros s/n, 39005 Santander, Spain}

\newcommand{\ISC}{Consiglio Nazionale delle Ricerche, Istituto dei Sistemi Complessi (CNR-ISC), via Madonna del Piano 17 50022 Sesto Fiorentino (FI), Italy
}

\newcommand{\INAF}{INAF-Osservatorio Astronomico di Arcetri, Largo Enrico Fermi 5 50125 Firenze Italy
}

\newcommand{\INFN}{
INFN-Sezione di Firenze, via G. Sansone 1 50022 Sesto Fiorentino, Italy
}

\begin{document}

%---- Title
\title{Sharpening the dark matter signature in gravitational waveforms II:\\Numerical simulations with the \texttt{NbodyIMRI} code}
\author{Bradley J. Kavanagh}
\email{kavanagh@ifca.unican.es}
\affiliation{\IFCA}
\author{Theophanes K. Karydas}
\email{tk.karydas@gmail.com}
\affiliation{\GRAPPA}
\author{\\Gianfranco Bertone}
\email{g.bertone@uva.nl}
\affiliation{\GRAPPA}

\author{Pierfrancesco Di Cintio}
\email{pierfrancesco.dicintio@cnr.it}
\affiliation{\ISC}
\affiliation{\INAF}
\affiliation{\INFN}

\author{Mario Pasquato}
\email{mario.pasquato@gmail.com}
\affiliation{D\'{e}partement de Physique, Universit\'{e} de Montr\'{e}al, 1375 Avenue Th\'{e}r\`{e}se-Lavoie-Roux, Montr\'{e}al, Canada}
\affiliation{Mila -- Quebec Artificial Intelligence Institute,
6666 Rue Saint-Urbain, Montr\'{e}al, Canada}
\affiliation{Ciela -- Montr\'{e}al Institute for Astrophysical Data Analysis and Machine Learning, Montr\'{e}al, Canada}
\affiliation{Dipartimento di Fisica e Astronomia, Universit\`{a} di Padova, Vicolo dell'Osservatorio 5, Padova, Italy}
\affiliation{Istituto Nazionale di Fisica Nucleare, Padova, Via Marzolo 8, Padova, Italy}

\begin{abstract}
Future gravitational wave observatories can probe dark matter by detecting the dephasing in the waveform of binary black hole mergers induced by dark matter overdensities. Such a detection hinges on the accurate modelling of the dynamical friction, induced by dark matter on the secondary compact object in intermediate and extreme mass ratio inspirals. In this paper, we introduce \texttt{NbodyIMRI}, a new publicly available code designed for simulating binary systems within cold dark matter `spikes'. Leveraging higher particle counts and finer timesteps, we validate the applicability of the standard dynamical friction formalism and provide an accurate determination of the maximum impact parameter of particles which can effectively scatter with a compact object, across various mass ratios. We also show that in addition to feedback due to dynamical friction, the dark matter also evolves through a `stirring' effect driven by the time-dependent potential of the binary. We introduce a simple semi-analytical scheme to account for this effect and demonstrate that including stirring tends to slow the rate of dark matter depletion and therefore enhances the impact of dark matter on the dynamics of the binary.

%Additionally, we introduce a simple semi-analytical scheme to evaluate and account for the effect of the dark matter stirring on the overall friction force acting on the secondary compact object. The latter is validated with ad hoc numerical calculations.
\end{abstract}

\keywords{Dark matter --- Gravitational waves --- Astronomical black holes --- Numerical simulations in gravitation and astrophysics}

\maketitle

%---------------------------------------------
%---------------------------------------------
\section{Introduction}

Gravitational waves (GWs) are a promising avenue for constraining and perhaps even detecting Dark Matter (DM)~\cite{Bertone:2018krk,Bertone:2019irm,Baryakhtar:2022hbu}. 
Future space-based GW observatories such as LISA \cite{LISA} and Taiji \cite{Hu:2017mde} will be sensitive to sub-Hertz GW frequencies, allowing them to follow the evolution of intermediate- and extreme-mass ratio
inspirals (IMRIs and EMRIs, respectively) over long times and providing a sensitive probe of black hole (BH) environments. 
Depending on the formation and growth history of the BH, this environment may include large overdensities of DM, known as DM spikes. 
 Potential hosts for DM spikes include astrophysical intermediate mass black holes (IMBHs) which have grown adiabatically at the centres of DM halos~\cite{Gondolo:1999ef,Ullio:2001fb,Bertone:2005hw} or Primordial Black Holes (PBHs), which form deep in the radiation era~\cite{Mack:2006gz,Eroshenko:2016yve,Adamek:2019gns,Green:2020jor}. 
 
The presence of DM spikes around massive BHs would impact the dynamics of inspiralling IMRIs and EMRIs, which would be imprinted in the phase evolution of GWs emitted by the system~\cite{Eda1,Eda2,Macedo:2013qea,Barausse:2014tra,Barausse:2014pra}; in the same way as central stellar cusps affect the dynamics around massive BHs in nuclear star clusters~\cite{2014ApJ...780..148A}. If observed over a large number of orbits (or equivalently a large number of GW cycles), this DM `dephasing' signature should be distinguishable from other environmental effects~\cite{Becker:2022wlo,Cole:2022yzw} and should allow for the detection of DM spikes and the extraction of information about the DM density profile~\cite{Cardoso:2019rou,Edwards:2019tzf,Coogan:2021uqv,Cole:2022ucw,Zhang:2024ugv}.

Though DM dephasing is a promising signature for future GW observatories, detecting dephased signals and correctly inferring their properties will require accurate waveforms. This in turn requires accurate modelling of the dynamics of binaries embedded in DM spikes. Initial studies of dephasing focused on quasi-circular Newtonian binaries, embedded in a static DM spike~\cite{Yue:2017iwc,Yue:2018vtk,Hannuksela:2019vip,Edwards:2019tzf}. A range of other effects have also been subsequently studied, including accretion~\cite{Yue:2017iwc,Nichols:2023ufs}; post-Newtonian corrections~\cite{Speeney:2022ryg,Montalvo:2024iwq}; and the impact of eccentric orbits~\cite{Yue:2019ozq,Tang:2020jhx,Li2,Becker:2021ivq,2023arXiv231214041H,Mukherjee:2023lzn,AbhishekChowdhuri:2023cle}. 

Dynamical friction (DF)~\cite{Chandrasekhar1943a,Chandrasekhar1943b, Chandrasekhar1943c,Dosopoulou:2023umg} is typically assumed to be the main mechanism by which the DM spike affects the binary. However, the standard DF formalism assumes a uniform, infinite medium, an assumption which is broken in the case of compact objects orbiting in a DM spike. Furthermore, the strength of the DF force is \textit{a priori} unknown, as it depends on the maximum impact parameter of particles which can effectively scatter with the compact object, $b_\mathrm{max}$. In addition, many previous studies have neglected the impact of feedback on the DM spike, which may reduce the dephasing signal by orders of magnitude~\cite{Kavanagh:2020cfn}. Though a formalism for including this feedback effect was presented in Kavanagh \textit{et al.}~\cite{Kavanagh:2020cfn}, it has yet to be verified. Detailed numerical study of the DF force in these systems (and the associated feedback) is therefore crucial. 

 In this paper, we present the $N$-body code \texttt{NbodyIMRI}~\cite{NbodyIMRI}, designed to allow fast and accurate simulation of binaries embedded in DM spikes. We present detailed numerical checks of the code and use it to estimate the strength of the dynamical friction force in DM spikes. We also use the code to study the feedback on the spike due to dynamical friction, demonstrating that previous semi-analytical approaches (such as \texttt{HaloFeedback}~\cite{HaloFeedback}) provide a good order-of-magnitude estimate for the depletion of the spike, but highlighting new effects which have not previously been considered. Reference~\cite{Mukherjee:2023lzn} recently presented $N$-body simulations of binaries in DM spikes, focusing on the evolution of eccentric orbits over long timescales. While our focus is on shorter timescales, we provide a brief comparison with our code and results in \cref{sec:Conclusions}.

In a companion paper~\cite{BetterSpikesI} (hereafter Paper I), we extend the formalism of feedback developed by Kavanagh \textit{et al.}~\cite{Kavanagh:2020cfn} to the case of eccentric orbits, as well as developing a formalism to account for dephasing and feedback due to accretion onto the orbiting companion. In addition, we use the \texttt{NbodyIMRI} code in Paper I to simulate and validate the accretion formalism developed there.

In \cref{sec:Framework}, we describe the simulation framework upon which the \NbodyIMRI{} code is based, as well as presenting numerical tests to ensure the stability of the simulations. In \cref{sec:Results}, we present measurements of the strength of the dynamical friction force in circular (\cref{sec:circular}) and eccentric (\cref{sec:eccentric}) orbits. In \cref{sec:Feedback}, we simulate the evolution of the DM spike with time. Finally, in \cref{sec:Conclusions} we compare our results with previous studies in the literature and present our conclusions.

%---------------------------------------------
%---------------------------------------------
\section{Simulation Framework}
\label{sec:Framework}

We begin by describing the $N$-body simulations we have developed, the code for which is publicly available.\footnote{\url{https://github.com/bradkav/NbodyIMRI}} These simulations describe the Newtonian dynamics of a central massive body $m_1$ surrounded by a DM spike, represented by psuedo-particles of mass $\tilde{m}_\mathrm{DM}$. A lighter companion $m_2$ orbits through the spike. 

\subsection{Force calculations}

\begin{figure}[tb]
    \begin{center}
        \includegraphics[width=0.80\linewidth]{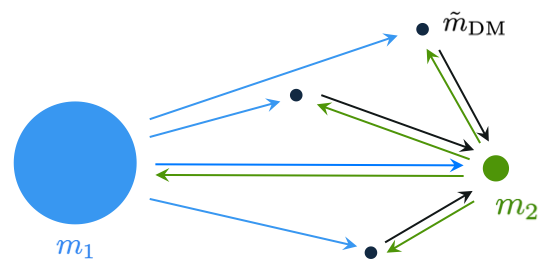}%
        \caption{\textbf{Schematic of the force calculations in \texttt{NbodyIMRI}.} The gravitational force on body $B$ due to body $A$ is indicated by an arrow ($A \rightarrow B$). The code includes unsoftened force calculations between the two BHs, $m_1$ and $m_2$. The force on the DM pseudo-particles $\tilde{m}_\mathrm{DM}$ due to the two BHs is softened, as described in the text. No pair-wise forces are calculated between DM particles, as the gravitational potential due to the DM spike is expected to be subdominant compared to that of the binary.}
    \label{fig:ForceDiagram}
    \end{center}
\end{figure}

 We assume that the dynamics of the DM particles within the spike will be dominated by the potential from the massive BHs. We will thus neglect pairwise force calculations between individual DM particles.\footnote{See Refs.~\cite{2020IAUS..351..532V,Mukherjee:2023lzn} for similar approaches.}
 In practice, the DM particles feel only the potential from the two BHs; the smaller BH feels the potential due to the DM particles and the central BH; and the central BH feels only the potential from the smaller BH. This scheme is illustrated in Fig.~\ref{fig:ForceDiagram}. With this approach, the force calculations scale as $N$, the number of DM particles, rather than $N^2$.\footnote{We note that neglecting the interaction between DM pseudo-particles in the $N$-body simulations prevents the formation of a trailing wake behind $m_2$ and a leading overdensity (see e.g.~\cite{1983A&A...117....9M} for the case of a massive object in a stellar system). The net contribution of said wake to the slowing down of the massive perturber is only relevant for `gaseous' systems~\cite{1999ApJ...513..252O}, but becomes negligible in the case of a pressureless DM fluid considered here.}  

The maximum acceleration which can be accurately resolved in the simulations is limited by the fixed time-step. When the Newtonian gravitational force is large, then, it may be necessary to implement force softening~\cite{Athanassoula:1999wz,Dehnen:2000nh}. This softening will alter the dynamics of the system, but will prevent spuriously large gravitational scattering.   

The forces between the two BHs are not softened; in our simulations the separation between the BHs is never small enough to require softening. However, the forces acting on the DM particles \textit{are} softened. The force from the central BH is softened with a uniform sphere kernel with radius $\epsilon_1$~\cite{1993A&A...270..561P}. That is, we replace the point-mass of the central BH with a uniform density sphere of the same mass $m_1$ and radius $\epsilon_1$. The gravitational force on the DM particles is then:
\begin{equation}
    \bm{F}_1(r) = - \frac{G m_1 \tilde{m}_\mathrm{DM}}{r^2} \hat{\bm{r}} \times \begin{dcases}
        r/\epsilon_1 & \quad \text{for } r < \epsilon_1 \,,\\
        1 & \quad \text{for } r \geq \epsilon_1 \,.\\
    \end{dcases}
\end{equation}
The position of the DM particles with respect to the central BH is $\bm{r} = \bm{r}_\mathrm{DM} - \bm{r}_1$. This choice of softening leads to a simple form for the potential which is continuous at $r = \epsilon_1$ (see \cref{sec:initialisation}).

The forces between the companion $m_2$ and the DM particles are also softened. In this case, we set the gravitational acceleration force to zero for radius $r < \epsilon_2$:
\begin{equation}
    \bm{F}_2(r) = - \frac{G m_2 \tilde{m}_\mathrm{DM}}{r^2} \hat{\bm{r}} \times \begin{dcases}
        0 & \quad \text{for } r < \epsilon_2\,,\\
        1 & \quad \text{for } r \geq \epsilon_2 \,,\\
    \end{dcases}
\end{equation}
where $\bm{r} = \bm{r}_\mathrm{DM} - \bm{r}_2$. Physically, this corresponds to assuming that all of the mass of the companion is concentrated on a thin spherical shell of radius $\epsilon_2$. This choice of softening allows us to estimate the maximum velocity of DM particles in the simulation, allowing us to choose a suitable value of the time step $\Delta t$, as described in \cref{app:RequiredTimestep}.

\subsection{Time stepping}

We use a leapfrog algorithm with a fixed timestep to update the positions and velocities of the particles in the simulation. Leapfrog algorithms are symplectic~\cite{1990CeMDA..50...59K} and therefore provide better stability over large numbers of orbits than, for example, higher order Runge-Kutta schemes not expressed in exponential form~\cite{2017JCoPh.338..567M}. The most well-known leapfrog algorithm is the second-order leapfrog, or `Verlet' method~\cite{Verlet1967}. In this case, a full timestep of size $\Delta t$ can be achieved via the following steps:
\begin{align}
\begin{split}
    \bm{x}_{i + 1/2} &= \bm{x}_{i} + \frac{1}{2} \mathbf{v}_i\,\Delta t\,,\\
    \mathbf{v}_{i + 1} &= \mathbf{v}_i +  \bm{a}(\bm{x}_{i + 1/2}) \,\Delta t\,,\\
    \bm{x}_{i + 1} &= \bm{x}_{i} + \frac{1}{2} \mathbf{v}_{i + 1}\,\Delta t\,.
\end{split}
\end{align}
We implement the 4th-order ``Position Extended Forest-Ruth Like" (PEFRL) leapfrog~\cite{1990PhyD...43..105F,1990PhLA..150..262Y,2002CoPhC.146..188O}. We use a single global timestep for all particles in the system. With this prescription, the system can be followed at high resolution for tens to thousands of orbits on a single core. For example, simulating $N_\mathrm{DM} = 256\mathrm{k}$ DM particles for a single orbit of the binary (with $10^4$ time-steps per orbit) takes approximately $3000$ seconds on a single core. 

The dynamics of all particles in the simulation is Newtonian. We do not include any Post-Newtonian (PN) corrections. This means that the simulations cannot be used to describe the long-time-scale evolution of the binary systems as they do not include any prescription for energy losses due to GW emission. However, the simulations are well-suited to calibrating the dynamical friction force and studying feedback on short timescales.

We note that a similar computational approach has been used independently by Mukherjee \textit{et al.}~\cite{Mukherjee:2023lzn} to study DM feedback in eccentric orbits. In that case, PN terms up to order 2.5 we included (including GW emission). We provide a brief comparison with our work in \cref{sec:Conclusions}.

\subsection{Initialization}
\label{sec:initialisation}

We initialise the BH binary in either circular or eccentric orbits, with an initial semi-major axis given by $a_i = 100\,\risco$.\footnote{We use the notation $\risco = 6 G m_1/c^2$ to denote the innermost stable circular orbit (ISCO) of the central BH. For reference, a BH mass of $m_1 = 1000 \,M_\odot$ would have an ISCO of $\risco \approx 2.87 \times 10^{-10}\,\mathrm{pc}$.} This is a typical separation at which the binaries we consider would enter into the frequency band of experiments like LISA~\cite{2005CQGra..22S.355B}. When calculating the initial velocity of the compact objects, we do not include the contribution of the DM spike itself to the enclosed mass. 

The DM density profile which we use in the simulations is given by:
\begin{equation}
\label{eq:rhodm}
    \rho_\mathrm{DM}(r) = \rho_6 \left(\frac{r}{r_6}\right)^{-\gamma_\mathrm{sp}} \left( 1 + \frac{r}{r_t}\right)^{-\alpha}\,,
\end{equation}
where $r_6 = 10^{-6}\,\mathrm{pc}$ is a reference radius; $\rho_6$ specifies the normalisation of the DM spike; and $\gamma_\mathrm{sp}$ is the inner density slope~\cite{Coogan:2021uqv}. The profile is softly truncated at a radius $r_t$, at which the logarithmic slope transitions from $\gamma_\mathrm{sp}$ to $\gamma_\mathrm{sp} + \alpha$. In all simulations, we fix $r_t = 2000 \, \risco$, and $\alpha = 2$. We also fix the slope $\gamma_\mathrm{sp} = 7/3 \approx 2.33$, as expected for the adiabatic growth of an astrophysical BH at the centre of an NFW halo~\cite{Gondolo:1999ef}.

We assume that the DM distribution is isotropic and spherically symmetric, being initially described by a distribution function $f(\mathcal{E}) \propto \mathrm{d}N/\mathrm{d}^3\bm{r}\mathrm{d}^3\mathbf{v}$ which depends only on the relative specific energy $\mathcal{E} = \Psi(r) - \frac{1}{2}v^2$. In the softened potential of the central BH, the relative potential is given by:
\begin{align}
    \Psi(r) = \frac{G m_1}{r} \times
    \begin{dcases}
    \frac{1}{2}\frac{r}{\epsilon_1}\left( 3 - \frac{r^2}{\epsilon_1^2}\right) & \text{for } r < \epsilon_1\,,\\
     r & \text{for } \epsilon_1 \geq 1\,.
    \end{dcases}
\end{align}
We do not include the contribution of the secondary compact object to the gravitational potential. We calculate $f(\mathcal{E})$ through the Eddington inversion procedure~\cite{BinneyAndTremaine,Lacroix:2018qqh}, numerically solving:
\begin{equation}
\label{eq:eddington}
f(\mathcal{E})=\frac{1}{\sqrt{8}\pi^2}\int_0^\mathcal{E}\frac{{\rm d}^2\rho_\mathrm{DM}}{{\rm d}\Psi^2}\frac{{\rm d}\Psi}{\sqrt{\mathcal{E}-\Psi}}\,.
\end{equation}
We draw the radius and velocity $(r, v)$ of each of the $N_\mathrm{DM}$ particles by inverse transform sampling. First, we draw the radius of the particle from the distribution $P(r) \propto 4\pi r^2 \rho_\mathrm{DM}(r)$. Then, we draw the speed of the particle from the velocity distribution at that radius 
$f(v) = 4 \pi v^2 f\left(\mathcal{E}(r, v)\right)$. The angular position of each particle and the direction of its velocity vector are chosen at random, uniformly on the sphere. 

In principle, using the Osipkov-Merritt extension of the Eddington formalism (see \cite{1979SvAL....5...42O,1985AJ.....90.1027M}), one could produce DM distribution function with a tunable degree of orbital velocity anisotropy. We have performed some tests on the influence of said anisotropy on the DF. These are presented in Appendix~\ref{OManiso}, where we find that increasing the fraction of radial orbits typically decreases the strength of the dynamical friction force. 

%PlotVelocityDistribution.py
\begin{figure}[tb]
    \begin{center}
    \includegraphics[width=0.95\linewidth]{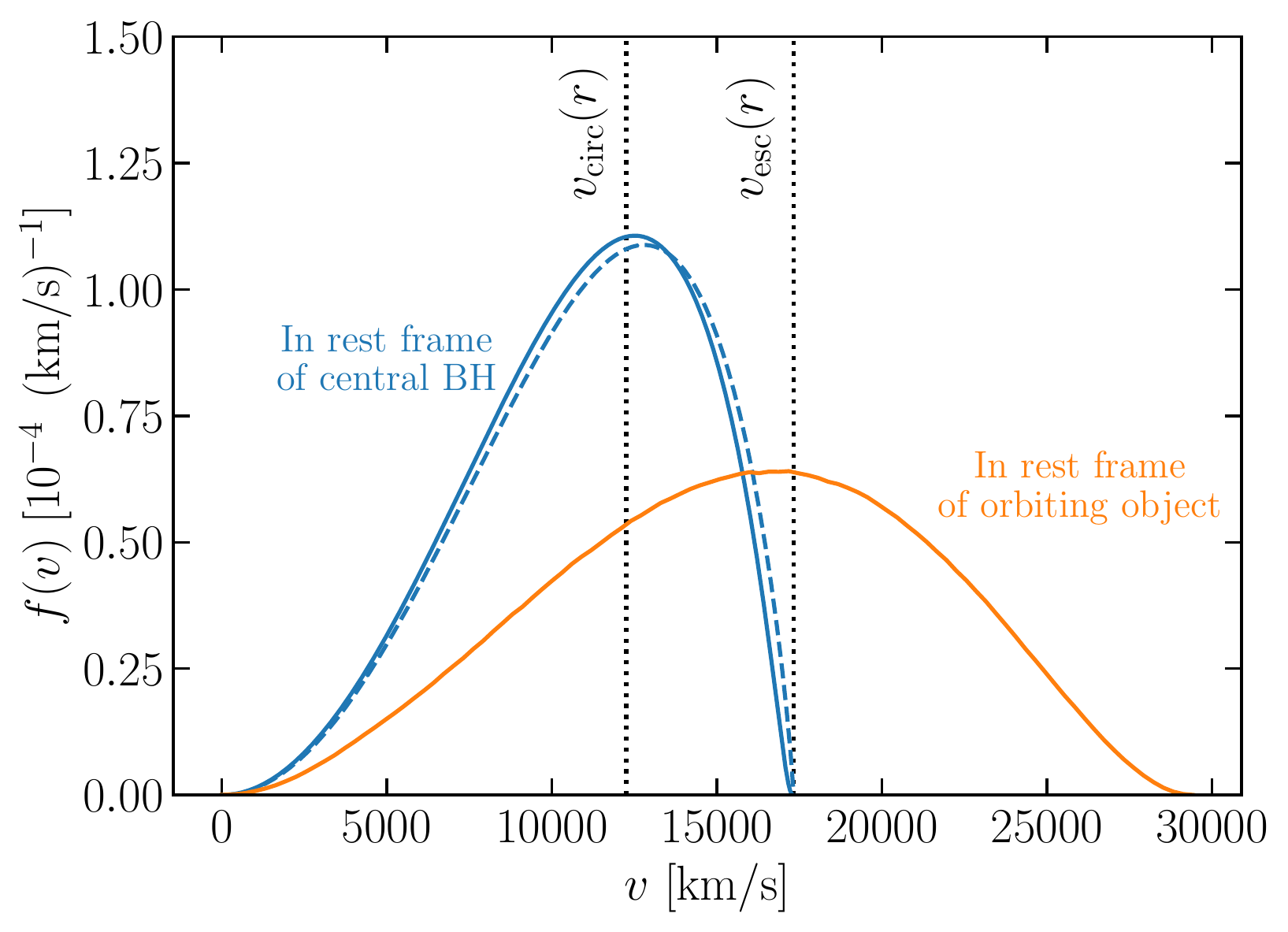}%
        \vspace{-0.2cm}
        \caption{\textbf{Velocity distribution of DM particles in the spike.} The blue curve shows the velocity distribution in the rest frame of the central black hole $m_1$, while the orange curve shows the velocity distribution relative to the orbiting compact object $m_2\ll m_1$. The results shown here are at a distance $r = 100 \,\risco$ from the central BH of mass $m_1 = 1000\,M_\odot$. We assume the orbiting object is on a circular orbit with velocity $v_\mathrm{circ}(r) = \sqrt{G m_1/r}$. In the rest frame of the central BH, the escape velocity of DM particles is $v_\mathrm{max}(r) =  \sqrt{2G m_1/r}$. The dashed blue line is the velocity distribution when the density profile is not truncated ($r_t \rightarrow \infty$).}
    \label{fig:VelocityDistribution}
    \end{center}
\end{figure}

The velocity distribution at $r = 100 \,\risco$ is illustrated in \cref{fig:VelocityDistribution}. The solid blue line shows the distribution in the rest frame of the central BH, where we have fixed the truncation radius of the profile to $r_t = 2000 \, \risco$ (as we use for the simulations in the rest of the paper). The truncation of the density profile has only a small impact on the velocity distribution (the blue dashed curve shows the distribution in the limit $r_t \rightarrow \infty$). The orange solid line shows the velocity distribution in the frame of a compact object orbiting at the local circular velocity.

\subsection{Numerical Tests}

Here, we describe two numerical tests to validate the long terms stability of the simulations. Unless otherwise stated, we assume $(m_1, m_2) = (1000, 1)\,M_\odot$.
%In Fig.~\ref{fig:BinaryStability}, we demonstrate the stability of a simulated binary in the absence of DM.
In order to check the reliability of the numerical scheme we integrated the dynamics of the binary in isolation, in the absence of a DM spike. Using $10^3$ timesteps per orbit, we find that the semi-major axis is stable up to $1$ part in $10^{13}$ while the eccentricity of the initially circular binary varies by, at most, $\Delta e \sim 3 \times 10^{-8}$. These variations are much smaller than the typical effect sizes we are hoping to detect, demonstrating the required stability of the simulations. Throughout the rest of this work, we will always use a minimum of $10^3$ timesteps per orbit. While this guarantees the stability of the binary, a shorter timestep may be required (depending on the simulation parameters) in order to resolve the dynamics of DM particles in close encounters with the secondary. More typically, then, we use $10^4$ timesteps, unless otherwise specified. 

%StabilityChecks.ipynb
%\begin{figure}[t]
%    \begin{center}
%        \includegraphics[width=0.95\linewidth]{figures/BinaryStability.pdf}%
%        \caption{\textbf{Stability of a vacuum binary during 1000 orbits.} We simulate a circular binary with masses $(m_1, m_2) = (1000, 1)\,M_\odot$ in the absence of a DM spike and show the change in semi-major axis (black, left) and the change in the eccentricity (orange, right). We use 1000 timesteps per orbit and $N_\mathrm{DM} = 256\mathrm{k}$ DM particles. \gfb{Not sure it make sense to plot the change in eccentricity here: there's no information in the plot beside what's written in the text $\Delta e \sim 3 \times 10^{-8}$, right? In fact, the plot only raises questions, I suspect..}}
%    \label{fig:BinaryStability}
%    \end{center}
%\end{figure}

%Analysis_Feedback.pdf
\begin{figure}[tb]
    \begin{center}
        \includegraphics[width=0.95\linewidth]
        {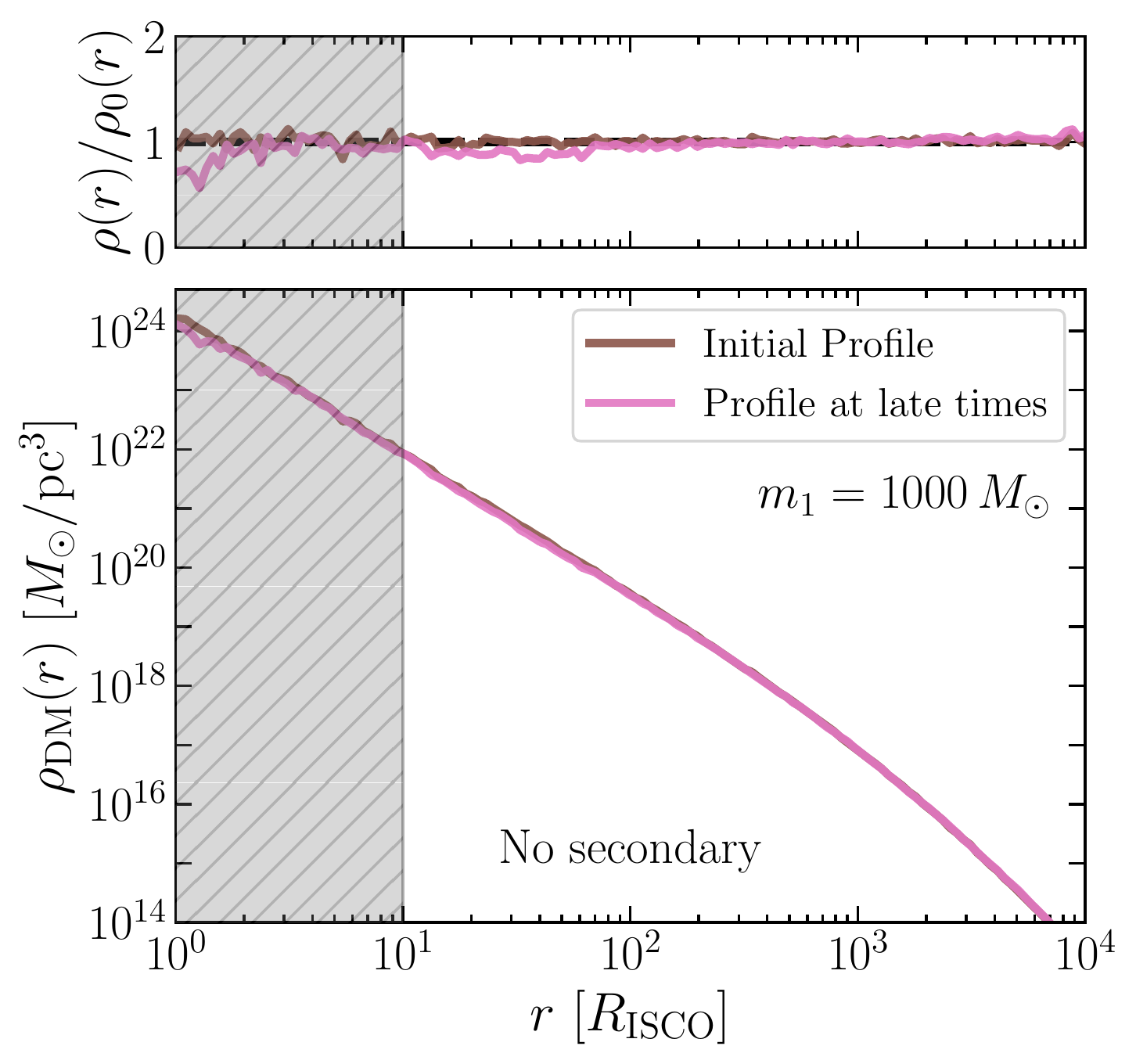}%
        \vspace{-0.2cm}
        \caption{\textbf{Stability of a simulated DM spike (in the absence of a secondary BH).} We simulate a central BH with mass $m_1 = 1000\,M_\odot$ surrounded by a DM spike. The interactions with the central BH are softened with $\epsilon_1 = 10\,\risco$ (grey shaded region). The simulation is run for the equivalent of 1000 orbits, for a binary with separation $a = 100\,\mathrm{isco}$.}
    \label{fig:ProfileStability}
    \end{center}
\end{figure}

In addition, we also performed a test simulation where $N_\mathrm{DM} = 256\mathrm{k}$ DM particles have been propagated in the potential of the central compact object without a companion. In Fig.~\ref{fig:ProfileStability}, we show the DM density profile of a simulated spike after a time $T$ in the absence of a secondary BH. This simulation covers a time equivalent to 1000 orbits of a binary at a separation $a_i = 100 \, \risco$ (though in this case the simulation consists only of a central BH $m_1$ surrounded by the spike).  We see that the density profile is stable at the level of a few percent at radii greater than $\sim 10\,\risco$. The interactions of the DM particles with the central BH are softened on scales smaller than $\epsilon_1 = 10\,\risco$, marked as a shaded grey region. However, the distribution function is initialised to reflect this softened potential, giving rise to good stability of the spike even in this softened region. The stability of the spike in this test simulation gives us confidence that changes in the spike with the inclusion of the secondary compact object are due to physical feedback effects, rather than poor initialisation and simulation of the system.

%---------------------------------------------
%---------------------------------------------
\section{Results}
\label{sec:Results}

In this section, we will present the results of simulations aimed at measuring the strength of the dynamical friction force due to the DM spike, as well as exploring how the DM spike responds to the energy injected by the orbiting companion. 

The dynamical friction force can be written as~\cite{BinneyAndTremaine}:
\begin{align}
\label{eq:DF}
    \bm{F}_\mathrm{DF} = -4 \pi \frac{(G m_2)^2}{u^2}\rho_\mathrm{DM}(r_2) \,\mathbf{\mathcal{C}}_\mathrm{DF}\hat{\bm{u}}\,,
\end{align}
where $\bm{u}$ is the orbital velocity of the companion, which orbits through the spike. Here, we have assumed that the medium is spherically symmetric and isotropic around the central BH, such that the DF acts as a drag force, pointing in the direction opposite to the object's motion.\footnote{In the most general case~\cite{BinneyAndTremaine}, the mass-dependent pre-factor in \cref{eq:DF} reads $m_2(m_2+\tilde{m}_\mathrm{DM})\rho_\mathrm{DM}$, where $\tilde{m}_\mathrm{DM}$ is the mass of the DM pseudo-particles in the $N$-body simulation. In our simulations where $m_\mathrm{DM}/m_2\approx 10^{-3}$ (and in general for particle DM), we can safely assume $m_2+\tilde{m}_\mathrm{DM}\simeq m_2$.} This force leads to an energy loss of the binary of $\dot{E}_\mathrm{DF} = -u F_\mathrm{DF}$, leading the binary to inspiral.  

The dimensionless \textit{dynamical friction coefficient} accounts for the relative velocities of the particles and the companion and can be written as:
\begin{align}
\label{eq:DF_coefficient}
    \mathcal{C}_\mathrm{DF} = -\int \ln \Lambda (\vrel) \frac{u^2}{\vrel^2}\left(\vrelhatvec \cdot \hat{\bm{u}}\right)\,f(\mathbf{v})\,\mathrm{d}^3\mathbf{v}\,.
\end{align}
Here, $\vrelvec = \mathbf{v} - \bm{u}$ is the velocity of the DM particles relative to the orbiting compact object
%\footnote{Note that the integral over $\mathbf{v}_\mathrm{rel}$ in Eq.~\eqref{eq:DF_coefficient} could just as well be replaced with an integral over the DM velocity distribution: $f(\mathbf{v}_\mathrm{rel})\,\mathrm{d}^3\mathbf{v}_\mathrm{rel} \rightarrow f(\mathbf{v})\,\mathrm{d}^3\mathbf{v}$.} 
and:
\begin{equation}
    \Lambda(\vrel) = \left[\frac{b_\mathrm{max}^2 + b_{90}(\vrel)^2}{b_\mathrm{min}^2 + b_{90}(\vrel)^2}\right]^{1/2}\,,
\end{equation}
is the Coulomb factor, such that $\ln \Lambda$ is the familiar Coulomb logarithm. The impact parameter for particles which contribute to the dynamical friction is assumed to lie in the range $b \in [b_\mathrm{min}, b_\mathrm{max}]$, while $b_{90}$ is the impact parameter for which an incoming particle undergoes a deflection of $90^\circ$:
\begin{align}
    b_{90}(\vrel) = \frac{G m_2}{\vrel^2}\,.
\end{align}

The dynamical friction coefficient is typically simplified by assuming that in the definition of $b_\mathrm{90}$ we can replace $\vrel$ by some typical velocity, such as $u$. In this way, the Coulomb logarithm can be removed from the integral. If the DM velocity distribution is isotropic, the integral further simplifies~\cite[Appendix L]{BinneyAndTremaine} to give:
\begin{align}
\begin{split}
\label{eq:DF_coefficient_simple}
    \mathcal{C}_\mathrm{DF} &= \ln \Lambda (u) \int_{v < u} f(\mathbf{v})\,\mathrm{d}^3\mathbf{v} \\
    &\equiv \ln \Lambda (u) \times \xi_{<u}\,.
\end{split}
\end{align}
This is the standard expression for the strength of the dynamical friction force, where we have used the notation $\xi_{<u}$ (similar to that in Ref.~\cite{Kavanagh:2020cfn}) to denote the fraction of DM particles moving more slowly that the orbital velocity $u$ of the compact object.
Note that, in a system with anisotropic $f(\mathbf{v})$ even particles moving {\it faster} than the in-spiraling compact object contribute to the DF force \cite{1977MNRAS.181..735B}. The same happens also in a relativistic set-up for isotropic models (e.g. see \cite{1994MNRAS.270..205S,2023A&A...677A.140C,Dosopoulou:2023umg}).

\subsection{Circular orbits}
\label{sec:circular}

In order to validate the expression for the dynamical friction force given in \cref{eq:DF,eq:DF_coefficient} and to estimate the value of the maximum impact parameter $b_\mathrm{max}$, we study the orbital decay of simulated binaries.

%Plot_Orbital_Evolution
\begin{figure}[tb]
    \begin{center}
    \includegraphics[width=0.95\linewidth]{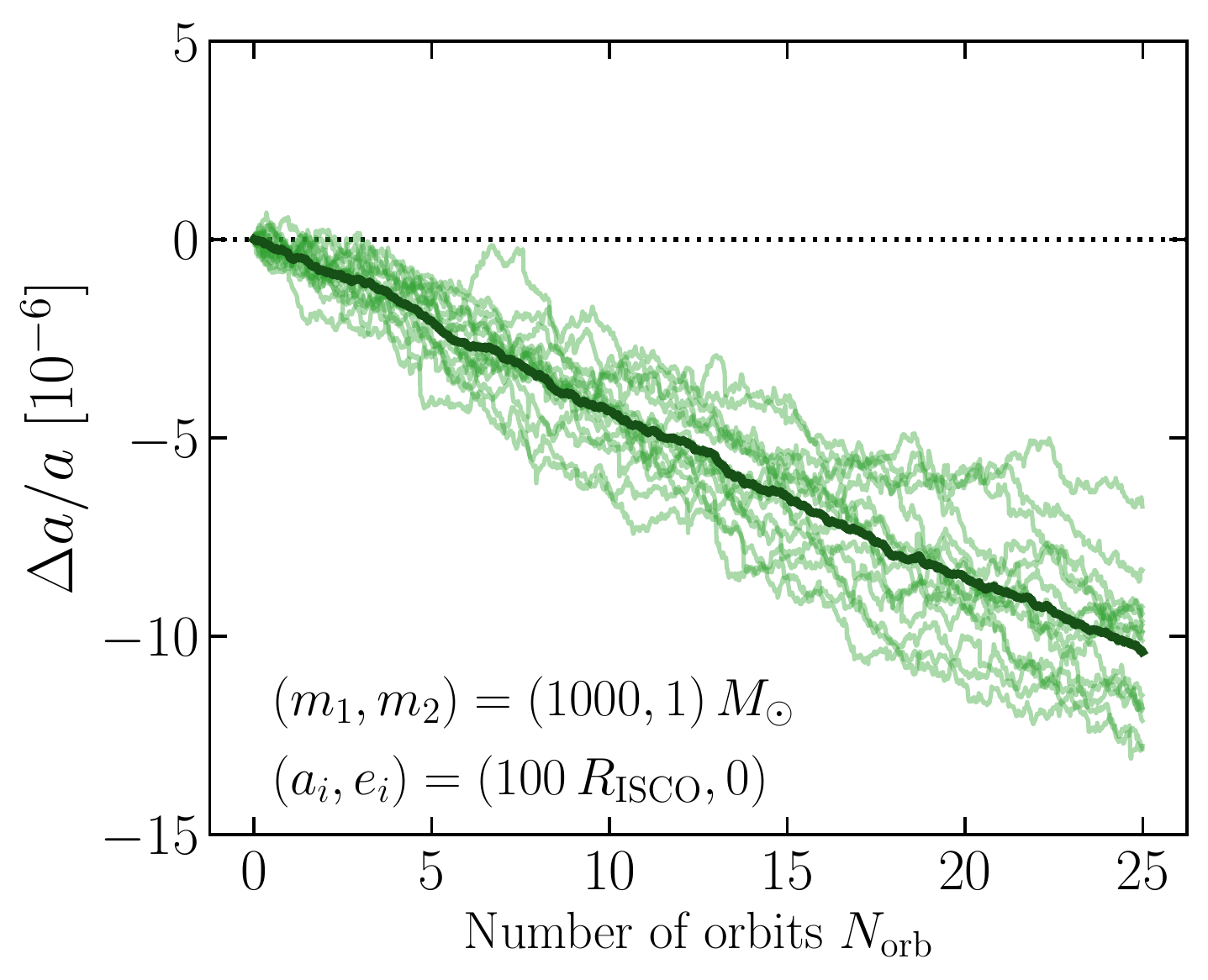}%
    \vspace{-0.2cm}
        \caption{\textbf{Decay of semi-major axis due to DM dynamical friction in $N$-body simulations.} Each of the 16 pale green curves shows the fractional change in the semi-major axis in a single realisation, simulating a circular binary with masses $(m_1, m_2) = (1000, 1)\,M_\odot$ in a DM spike, with a DM softening length of $\epsilon_2 = 10^{-3} a_i$. The thick curve shows the average over the ensemble of realisations. The decay in the semi-major axis is driven by dynamical friction with the DM spike, and the rate of decay allows us to infer the strength of the dynamical friction force, characterised by $\mathcal{C}_\mathrm{DF}$.}
    \label{fig:SemimajorAxis}
    \end{center}
\end{figure}

In \cref{fig:SemimajorAxis}, we plot the fractional change in semi-major axis of an initially circular binary with masses $(m_1, m_2) = (1000, 1)\,M_\odot$. We assume a binary separation of $a_i = 100\,\risco$ DM softening length of \linebreak $\epsilon_2 = 10^{-3} a_i$. The thick green line shows the average over an ensemble of 16 realisations of the DM spike, illustrating that over 25 orbits, the change in the semi-major axis is roughly 1 part in $10^5$. Variations between the different realisations arise from the discreteness of the DM distribution and can be suppressed by increasing the number of DM pseudo-particles. Here, we use $N_\mathrm{DM} = 256\,\mathrm{k}$ particles in order to resolve such small changes in the orbital parameters. The change in the semi-major axis can be used to estimate the rate of energy loss by dynamical friction; the orbital energy of the binary is given by $E_\mathrm{orb} = -G (m_1 + m_2)/(2 a)$, such that $\Delta a/a = \Delta E_\mathrm{DF}/E_\mathrm{orb}$. In this way, we can estimate the value of $\mathcal{C}_\mathrm{DF}$ from this ensemble of simulations.

\begin{figure*}[ht!]
    \begin{center}
        \includegraphics[width=0.95\linewidth]{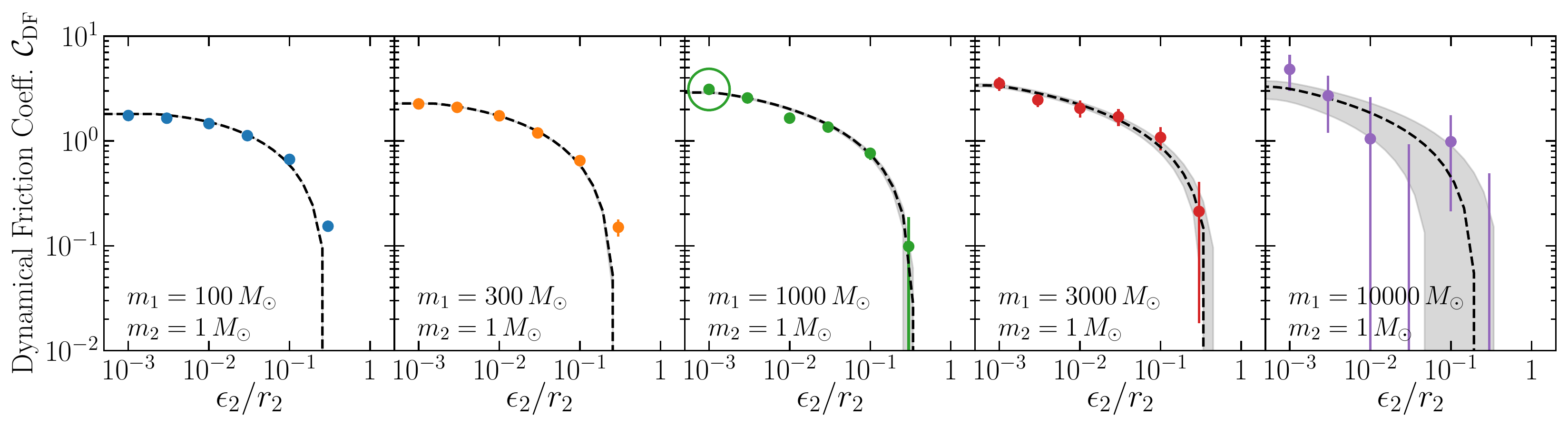}\\
\vspace{-0.16cm}\hspace{0.29cm}\includegraphics[width=0.941\linewidth]{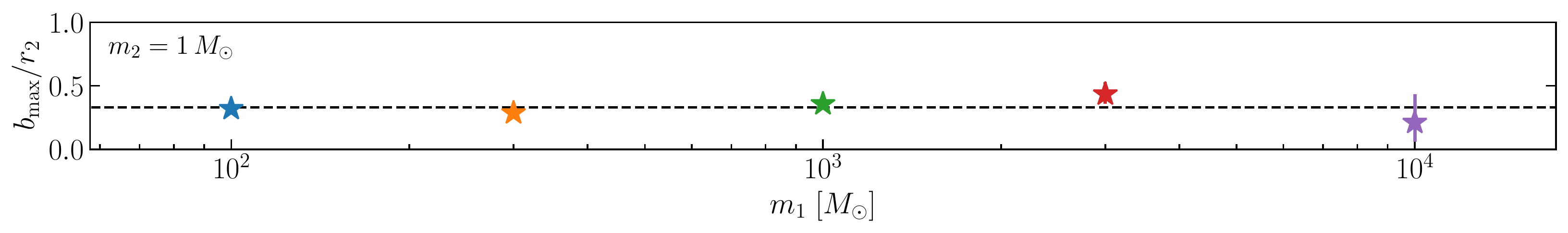} \vspace{-0.2cm}
        \caption{\textbf{Dynamical friction coefficient $\mathcal{C}_\mathrm{DF}$ estimated from $N$-body simulations.} Each point in the upper panels shows the DF coefficient estimated from the orbital decay $\Delta a/a$ of an ensemble of simulated binaries, assuming a specific value of the softening length of the secondary object $\epsilon_2$. We follow each binary for 25 orbits, simulating the DM spike with $N_\mathrm{DM} = 256\,\mathrm{k}$ particles. The point highlighted with a circle in the top-centre panel corresponds to the set of simulations shown in \cref{fig:SemimajorAxis}.
        Each of the stars in the lower panel shows the value of $b_\mathrm{max}$ obtained from a fit to the DF coefficients in the panel above for the corresponding mass. The dashed black lines in the upper panels show the values of $\mathcal{C}_\mathrm{DF}$ calculated from \cref{eq:DF_coefficient} for these best fitting $b_\mathrm{max}$ values (with errors shown as grey shaded regions). The horizontal dashed line in the lower panel shows the mean value across different masses: $b_\mathrm{max} = 0.3 \,r_2$.}
    \label{fig:DFcoefficient}
    \end{center}
\end{figure*}

In the top panel of \cref{fig:DFcoefficient}, we show the value of $\mathcal{C}_\mathrm{DF}$ inferred from ensembles of simulations with different central masses $m_1$, as a function of the softening length of the companion $\epsilon_2$.\footnote{We simulate using $3 \times 10^4$ timesteps per binary orbit for the $m_1 = 100\,M_\odot$ case and $10^4$ timesteps per orbit for the remaining masses.} The point marked by the green ring in the central panel corresponds to the ensemble of simulations shown in \cref{fig:SemimajorAxis}. For $m_1 = \{100,\,300,\,1000\}\,M_\odot$, each point corresponds to an ensemble of 16 simulations, while for $m_1 = \{3000,\,10000\}\,M_\odot$, we use 32 simulations. This is because as the central mass increases, the size of the dynamical friction effect (at $r = 100\,\risco$) decreases, making the orbital decay harder to resolve over a fixed number of orbits.

For a fixed $m_1$, we see from \cref{fig:DFcoefficient} that increasing the softening length decreases the strength of the dynamical friction force. This is because the softening length effectively sets the minimum impact parameter  for which DM particles can undergo large deflections with the secondary object, $b_\mathrm{min} \sim \epsilon_2$ (see e.g.~Ref.~\cite{2019JMP....60e2901C}). At low values of the softening length, the DF coefficient tends to a constant as $b_\mathrm{min}$ becomes negligible compared to $b_\mathrm{max}$ and $b_{90}$ and the Coulomb factor tends to $\Lambda \rightarrow \sqrt{b_\mathrm{max}^2 + b_{90}^2}/b_{90}$. However, at sufficiently large values, the softening length may become comparable to (or larger than) the maximum impact parameter, leading the DF force to cut off. 

For each mass $m_1$, we perform a $\chi^2$ fit of $\mathcal{C}_\mathrm{DF}(b_\mathrm{max})$ in order to estimate the value of $b_\mathrm{max}$. We evaluate the DF coefficient according to \cref{eq:DF_coefficient}, calculating the integral numerically over the simulated velocity distribution (as shown in \cref{fig:VelocityDistribution}) and setting $b_\mathrm{min} = \epsilon_2$. Each panel in the upper part of \cref{fig:DFcoefficient} provides an estimate of $b_\mathrm{max}$ which is shown in the lower panel. 

From this ensemble of simulations, we infer that the maximum impact parameter is consistent across at least 2 orders of magnitude in the mass ratio and is well fit by the value:
\begin{equation}
    b_\mathrm{max} \approx (0.31 \pm 0.04)\,r_2\,,
\end{equation}
where $r_2$ is the separation of the binary. We have verified that this estimate of $b_\mathrm{max}$ is robust to different initial separations of the binary and different values of $m_2$. In \cref{sec:Conclusions}, we compare these results with those of previous simulations in the literature.

%$\mathcal{C}_\mathrm{DF} = 3.1 \pm 0.1$ for $m_1 = 1000\,M_\odot$.

In addition to estimating $b_\mathrm{max}$, these simulations allow us to check the simplified DF formula given by \cref{eq:DF_coefficient_simple} For a circular orbit, setting $\vrel \approx v_\mathrm{circ}$, we have $b_{90} \approx (m_2/m_1) r_2$. Using $b_\mathrm{max} = 0.3\,r_2$, this gives $\ln\Lambda = 5.36$ for $m_1 = 1000\,M_\odot$. From the velocity distribution in our simulations, we can estimate the fraction of particles moving slower than the orbital speed, giving $\xi_{<u} \approx 0.61$. With this, we find  $\ln\Lambda \times \xi_{<u} = 3.5$, compared to the value $\mathcal{C}_\mathrm{DF} = 3.1 \pm 0.1$ which we find for simulations of $m_1 = 1000\,M_\odot$ for small softening lengths. 

Similar estimates for other values of $m_1$ suggest that the simplified expression for the DF coefficient given in \cref{eq:DF_coefficient_simple} may overestimate the DF force by $10$-$20\%$, though this conclusion may change for different velocity distributions. In contrast, the full calculation of $\mathcal{C}_\mathrm{DF}$ given in \cref{eq:DF_coefficient} (illustrated as black dashed lines in the upper panels of \cref{fig:DFcoefficient}) provides a good fit to the simulation results.

\subsection{Eccentric orbits}
\label{sec:eccentric}

The dynamical friction force due to a DM spike in eccentric orbits has been studied in a number of previous works~\cite{Yue:2019ozq,Tang:2020jhx,Li2,Becker:2021ivq}, as well as in Paper I~\cite{BetterSpikesI}. Here, we validate the strength of the DF force in eccentric orbits using simulations. This is particularly important because the separation of the binary can vary drastically over the course of a single orbit, meaning that the choice of $b_\mathrm{max}$ may be non-trivial, as it is typically related to this separation.

\begin{figure}[tb]
    \hspace*{-0.4cm}
    \includegraphics[width=\columnwidth]{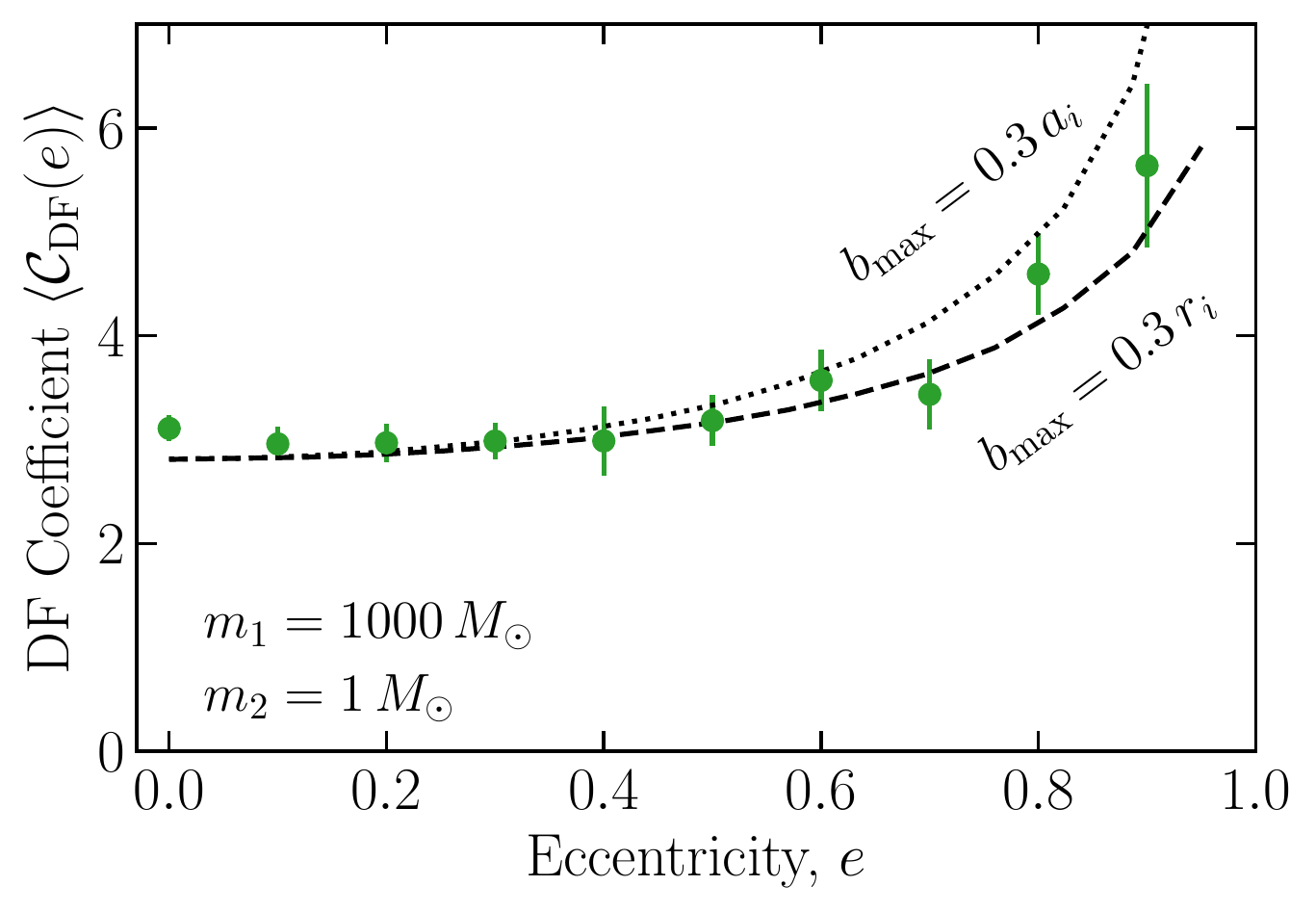}
    \vspace{-0.2cm}
    \caption{\textbf{Dynamical Friction coefficient for eccentric orbits.} We measure the strength of the dynamical friction force from the change in semi-major axis $\Delta a/a$. Here, $\left\langle\mathcal{C}_\mathrm{DF}(e)\right\rangle$ is the enhancement in the dynamical friction energy loss with respect to a circular orbit with the same semi-major axis, assuming a cloud of DM particles at rest. The dashed and dotted lines show the analytic estimates of $\left\langle\mathcal{C}_\mathrm{DF}(e)\right\rangle$ assuming $b_\mathrm{max} = 0.3 \,r_2$ and $b_\mathrm{max} = 0.3 \,a_i$ respectively.}
    \label{fig:sim_eccentricity}
\end{figure}

In \cref{fig:sim_eccentricity}, we plot the orbit-averaged DF coefficient $\left\langle \mathcal{C}_\mathrm{DF}(e)\right\rangle$ estimated from simulations. Each blue data point is obtained, as in the previous section, from 16 simulations of 25 orbits each. We define $\left\langle \mathcal{C}_\mathrm{DF}(e)\right\rangle$ as the value of the DF coefficient which would give rise to the same average energy loss per orbit as for a circular orbit with the same semi-major axis:
\begin{align}
     \left\langle \dot{E}_\mathrm{DF}\right\rangle = -4\pi   \frac{(G m_2)^2}{v_\mathrm{circ}}\rho_\mathrm{DM}(r = a_i) \left\langle \mathcal{C}_\mathrm{DF}(e)\right\rangle \,.
\end{align}
With this definition, then, the orbit-averaged DF coefficient tends to the value for circular orbits as $e \rightarrow 0$: $\left\langle \mathcal{C}_\mathrm{DF}(e = 0)\right\rangle \equiv \mathcal{C}_\mathrm{DF}$.
The black dashed line in \cref{fig:sim_eccentricity} shows the semi-analytic estimate of $\left\langle \mathcal{C}_\mathrm{DF}(e)\right\rangle$ obtained by explicitly orbit-averaging~\cite{Maggiore:2007ulw} the DF energy loss as calculated in \cref{eq:DF_coefficient}:
\begin{align}
  \label{eq:orbit_average}
  \left\langle \dot{E}_\mathrm{DF}\right\rangle = (1 - e^2)^{3/2} \int \displaylimits_0^{2\pi} \dot{E}_\mathrm{DF}(\theta)\, (1 +e\cos\theta)^{-2}\, \frac{\mathrm{d}\theta}{2\pi}\,.
\end{align}
Here, $\theta$ is the true anomaly, specifying the angular position of the compact object along the eccentric orbit. The energy loss $\dot{E}_\mathrm{DF}$ depends on $\theta$ as the DM density, the orbital velocity, and the DM velocity distribution vary along the orbit.

%Estimate_bmax.ipynb

We see from \cref{fig:sim_eccentricity} a good agreement between the DF coefficient estimated from simulations and that obtained by orbit-averaging. Calculating the DF force with $b_\mathrm{max} = 0.3\,r_2$ (black dashed line) provides a good fit to the data $\chi^2/\mathrm{dof} \approx 1.0$. There is no  strong preference for this over the alternative choice $b_\mathrm{max} = 0.3\,a_i$ (black dotted line, $\chi^2/\mathrm{dof} \approx 1.5$). However, the latter appears to slightly over estimate the DF force, especially at large eccentricities.

The behaviour of \cref{fig:sim_eccentricity} can be understood straightforwardly: DF coefficient increases with eccentricity as the energy loss is dominated by parts of the orbit where the compact object is slow-moving at large radii. The system spends a larger time at these large radii, and the reduced density at large $r$ is overcome by the scaling of the energy loss with $1/u$. This is an important validation of the dynamical friction formalism in eccentric systems and the validation of the value of $b_\mathrm{max}$ provides information on how to implement feedback in such systems. In Paper I, we describe an analytic prescription for feedback on the DM spike in eccentric orbits.

\subsection{Spike Feedback}
\label{sec:Feedback}

We now study the impact of the motion of the binary on the distribution of DM particles. We simulate a binary with a relatively close mass ratio, specifically a circular binary with $(m_1, m_2) = (100, \,1)\,M_\odot$. As we increase the central mass $m_1$, the DM spike particles become more tightly bound, and as we decrease the mass of the companion $m_2$, the energy injected via dynamical friction decreases. We therefore expect the depletion timescale for the DM spike to scale as $t_\mathrm{dep} \sim (m_1/m_2)^2 T_\mathrm{orb}$~\cite{Coogan:2021uqv}. Fixing $m_1/m_2 = 100$, we expect a depletion timescale on the order of thousands of orbits. Using the current simulation setup, larger mass ratio systems would require prohibitively long simulation times before substantial feedback in the spike is observed.  We fix the density normalisation of the spike to $\rho_6 = 9.1 \times 10^{14} \,M_\odot/\mathrm{pc}^3$ and simulate a binary with initial separation $a_i = 100\,\risco$.\footnote{The value of $\rho_6 = 9.1 \times 10^{14} \,M_\odot/\mathrm{pc}^3$ corresponds to $\rho_\mathrm{sp} = 226\,M_\odot/\mathrm{pc}^3$~\cite{Coogan:2021uqv}. This simulated system is comparable to the one considered in Appendix A of Ref.~\cite{Mukherjee:2023lzn}.} We simulate each system using $N_\mathrm{step} = 3 \times 10^4$ timesteps per orbit and a softening length of $\epsilon_2 = 0.001\,a_i$.

\begin{figure}[t]
    \begin{center}
        \includegraphics[width=0.45\textwidth]{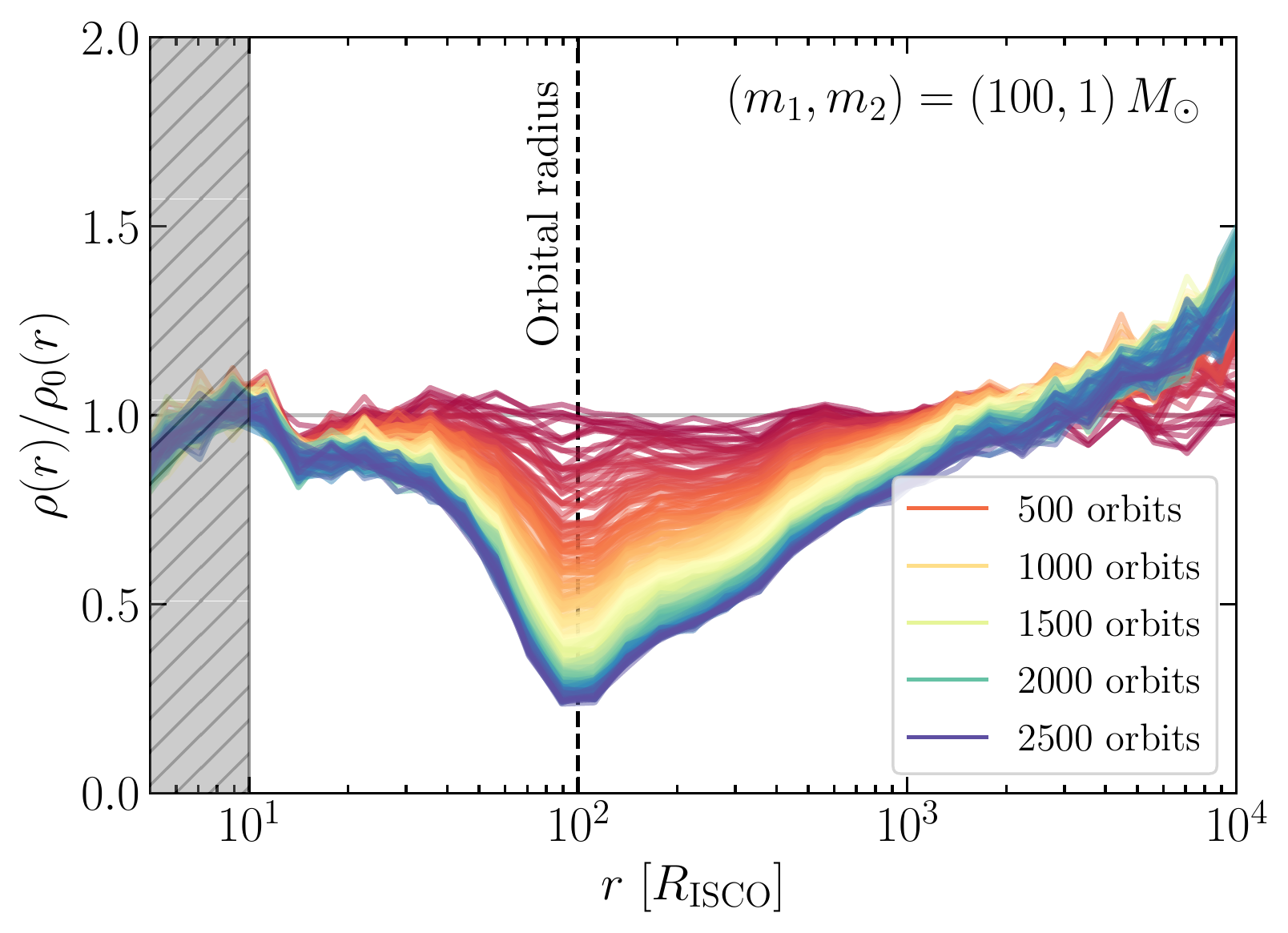}\\
        \includegraphics[width=0.45\textwidth]{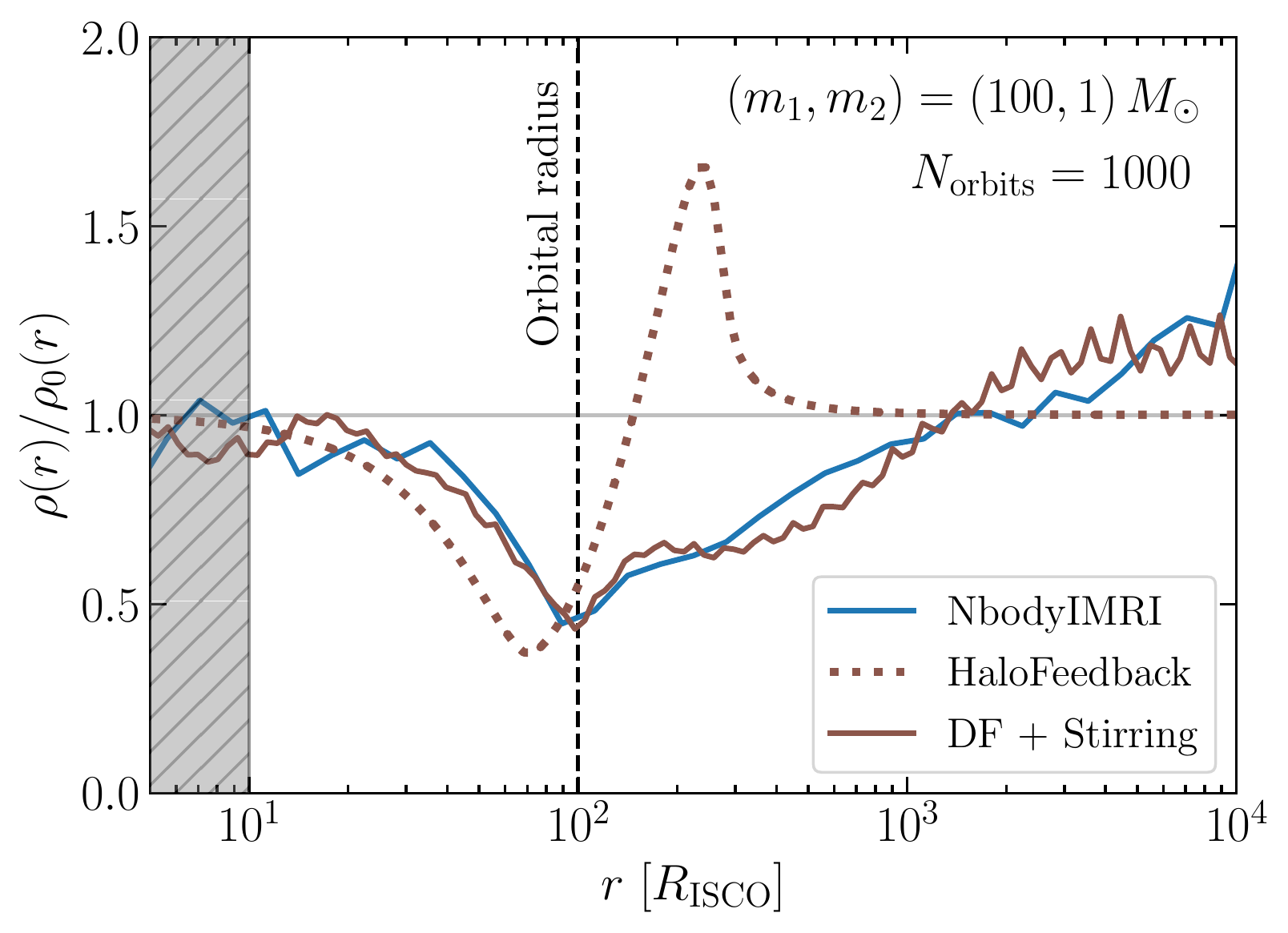}%
        \caption{\textbf{Evolution of the DM density profile during the simulation.} The upper panel shows the ratio of the final and initial DM density profiles for a circular system with $(m_1, m_2) = (100, 1)\,M_\odot$ over 2500 orbits. The lower panel shows a snapshot of the density profile after $1000$ orbits. The blue line shows the results of the \NbodyIMRI{} simulations, while the dotted brown line shows the density profile predicted by \texttt{HaloFeedback}~\cite{HaloFeedback}. The solid brown line shows the profile predicted by the feedback mechanism outlined in the text, including both close encounters with the secondary (``DF") and the energy injected by the time-varying potential of the binary (``Stirring").}
    \label{fig:Feedback}
    \end{center}
\end{figure}

In the upper panel of \cref{fig:Feedback}, we show the ratio of the initial and final density in the DM spike over 2500 orbits in the \NbodyIMRI{} simulations. We simulate 16 independent DM spike realisations, each comprising $N_\mathrm{DM} = 1\,\mathrm{k}$, and reconstruct the DM density profile by averaging over realisations (and averaging over every 10 orbits of the binary).  The most substantial decrease in the DM density is seen at the position of the orbital radius (vertical dashed line), where the DM density drops to around 25\% of its initial value. The particles depleted from this region are pushed to larger radii, as seen by the increased density above $r = 10^3 \,\risco$.

In the lower panel of \cref{fig:Feedback}, we compare the DM density profile obtained after 1000 orbits using the \NbodyIMRI{} code (solid blue line) with the results of the \texttt{HaloFeedback} code~\cite{HaloFeedback} (dotted brown line). \texttt{HaloFeedback} was introduced in Ref.~\cite{Kavanagh:2020cfn}, using a semi-analytical description to evolve the DM density profile as energy is injected into the spike through dynamical friction. We see that while the \texttt{HaloFeedback} formalism provides a good order-of-magnitude estimate for the density profile close to the orbital radius, there is a substantial discrepancy at large radii. \texttt{HaloFeedback} predicts a narrow overdensity in the DM density profile at a few times the orbital radius, while our simulations show a smoothly depleted region out to radii $\sim 10$ times larger. 

\begin{figure}[t]
    \begin{center}
        \includegraphics[width=0.49\textwidth]{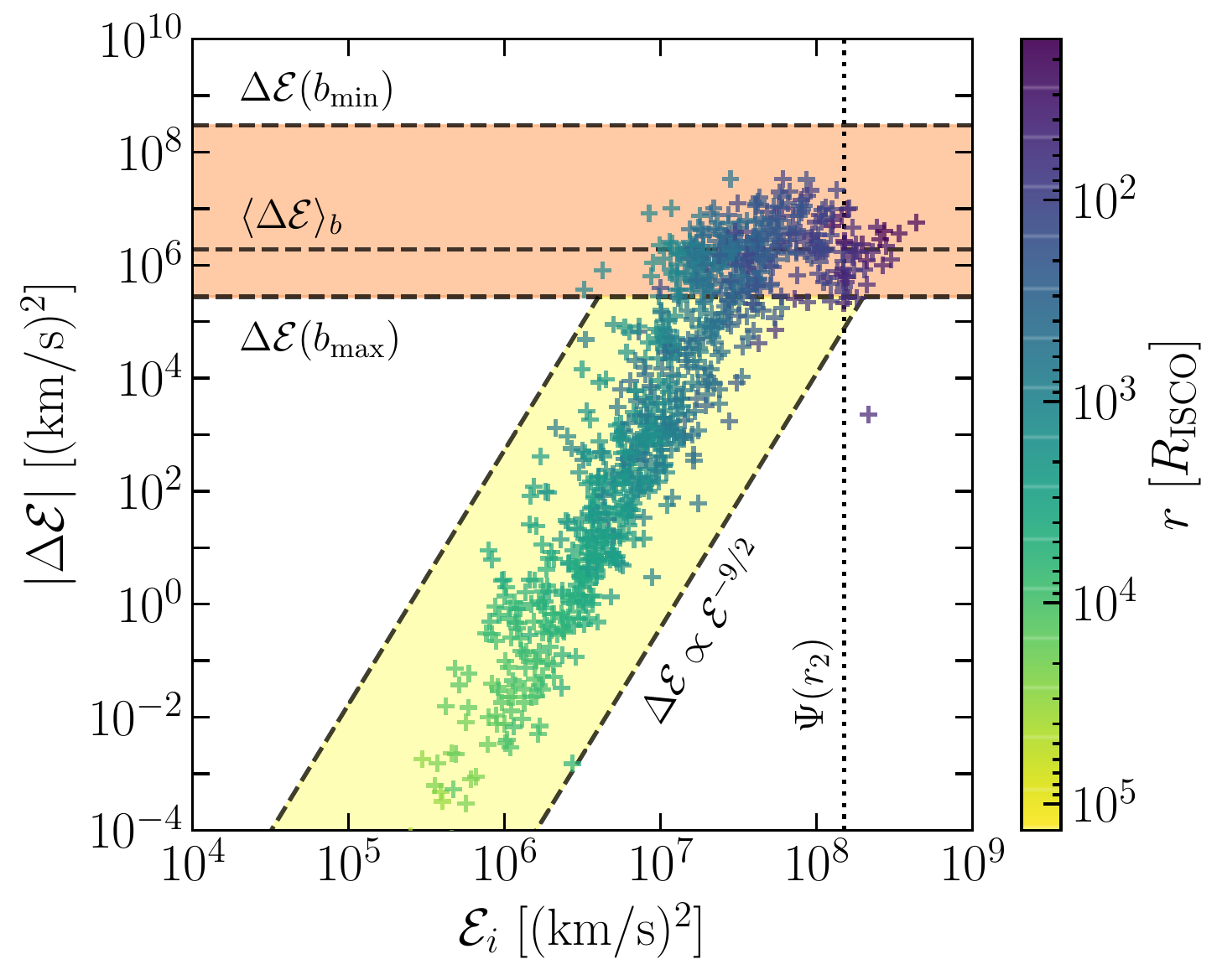}%
        \vspace{-0.2cm}
        \caption{\textbf{Change in energy $\Delta\mathcal{E}$ of DM pseudo-particles after 10 binary orbits.} Each point corresponds to a single DM pseudo-particle, colored according to its initial distance from the central BH (the secondary compact object is orbiting at a distance $r_2 = 100\,\risco$). The typical values of $\Delta \mathcal{E}$ expected from close encounters with the secondary object are shown as a shaded orange region. In addition, there is broad distribution of particles at large radii with $\Delta \mathcal{E} \propto \mathcal{E}^{-9/2}$ which cannot be explained by close two-body encounters. The vertical dotted line denotes the potential at the orbital radius of the secondary $\Psi(r_2)$.}
    \label{fig:DeltaE}
    \end{center}
\end{figure}

In order to investigate the origin of this discrepancy, in \cref{fig:DeltaE} we show the change in energy $\Delta\mathcal{E}$ for particles in the DM spike after 10 simulated orbits of the binary. Dynamical friction arises from two-body scattering between DM particles and the secondary. This two-body scattering is expected to lead to changes in the energy lying in the orange shaded region, where:
\begin{equation}
\label{eq:DeltaE_DF}
    |\Delta \mathcal{E}| \approx 2 u^2\left[1 + \frac{b^2}{b_{90}^2}\right]^{-1} \,.
\end{equation}
This is the change in energy implemented in the \texttt{HaloFeedback} formalism, with $b_\mathrm{min} = \epsilon_2$ and $b_\mathrm{max} = 0.3\,r_2$ (as described in \cref{sec:circular}).

We see from \cref{fig:DeltaE} that there is a large population of DM particles, particularly those at large radii, which experience a small change in energy, outside that expected for two-body close encounters. We find that these particles experience a change in energy which scales as $\Delta \mathcal{E} \propto \mathcal{E}^{-9/2}$, as highlighted by the yellow shaded region. We emphasise that these particles are at radii much larger than the binary separation, meaning that close encounters with the binary cannot account for this change in energy. 

%PlotDeltaE.ipynb
%DeltaE_plot_7Ece4.hdf5

This additional energy injection can be understood in terms of the time-dependent potential generated by the binary. In a static potential, the energy $\mathcal{E}$ of the DM particles is conserved. However, the orbiting binary generates a time-dependent potential which over a time $\Delta t$ will lead to a change in the DM energy given by:
\begin{equation}
    \Delta \mathcal{E} = \int_0^{\Delta t} \frac{\partial \psi}{\partial t}\,\mathrm{d}t\,.
\end{equation}
Here, the integral is taken following the path of the DM particle. A detailed calculation is presented in \cref{app:Stirring}, but we find that over a single orbit of a circular binary, this change in energy can be written as:
\begin{align}
\label{eq:DeltaE_stirring}
    \Delta \mathcal{E} = \left(\frac{G m_2}{r}\frac{a^2}{r^2}\right)\left(\frac{v T_\mathrm{orb}}{r}\right) \mathcal{A}(\hat{\bm{r}},\hat{\mathbf{v}})\,,
\end{align}
where $r$ is the radial position of the DM particle and $v$ is its velocity. The term in the first set of brackets corresponds to the time-dependent component of the potential due to the binary (assuming $r \gg a$).\footnote{The gravitational monopole and dipole moments are constant with time, so the leading order time-dependent contribution is from the quadrupole, whose potential scales as $a^2/r^3$.} The second term in brackets corresponds to the fractional change in the position of the particle over a single orbit. The factor $\mathcal{A}(\hat{\bm{r}},\hat{\mathbf{v}})$ contains the information on the angular position and velocity of the particle, see \cref{eq:dE_stirring}. Note that this mechanism may increase or decrease the energy of DM particles, depending on the sign of $\mathcal{A}(\hat{\bm{r}},\hat{\mathbf{v}})$.\footnote{We show in \cref{app:Stirring} that for a circular orbit and a spherically symmetric DM distribution, this stirring effect should lead to no net enegry loss of the binary.} The typical radial position of a particle scales as $r \sim G m_1/\mathcal{E}$, while the typical velocity scales as $v \sim \sqrt{\psi} \sim \sqrt{\mathcal{E}}$. The change in energy due to this `stirring' effect thus scales as $\Delta \mathcal{E} \propto \mathcal{E}^{-9/2}$ as shown in \cref{fig:DeltaE}.

In order to determine whether this `stirring' can explain the discrepancy between the results of \NbodyIMRI{} and those of \texttt{HaloFeedback}, we perform a toy calculation of the feedback. We generate sample of 10000 DM particles from the initial spike distribution and keep track of their relative specific energies $\mathcal{E}$. For each orbit of the binary, we calculate the probability of scattering with the companion and, in the event of an encounter, we modify the particle's energy according to \cref{eq:DeltaE_DF}.\footnote{See Ref.~\cite{Kavanagh:2020cfn} or Paper I for further details on the calculation of these scattering probabilities.} For each orbit, we also draw the radius and velocity of each particle and modify its energy according to \cref{eq:DeltaE_stirring} (as long as the particle lies outside the binary, $r > a$). We then reconstruct the DM density profile at each step from the distribution of energies $\mathcal{E}$. While a more complete description of this `stirring' effect should follow the energy injected in individual particle trajectories over long timescales, this simple toy model allows us to estimate the effect on the overall DM distribution.

\begin{figure}[t]
    \begin{center}
\includegraphics[width=0.45\textwidth]{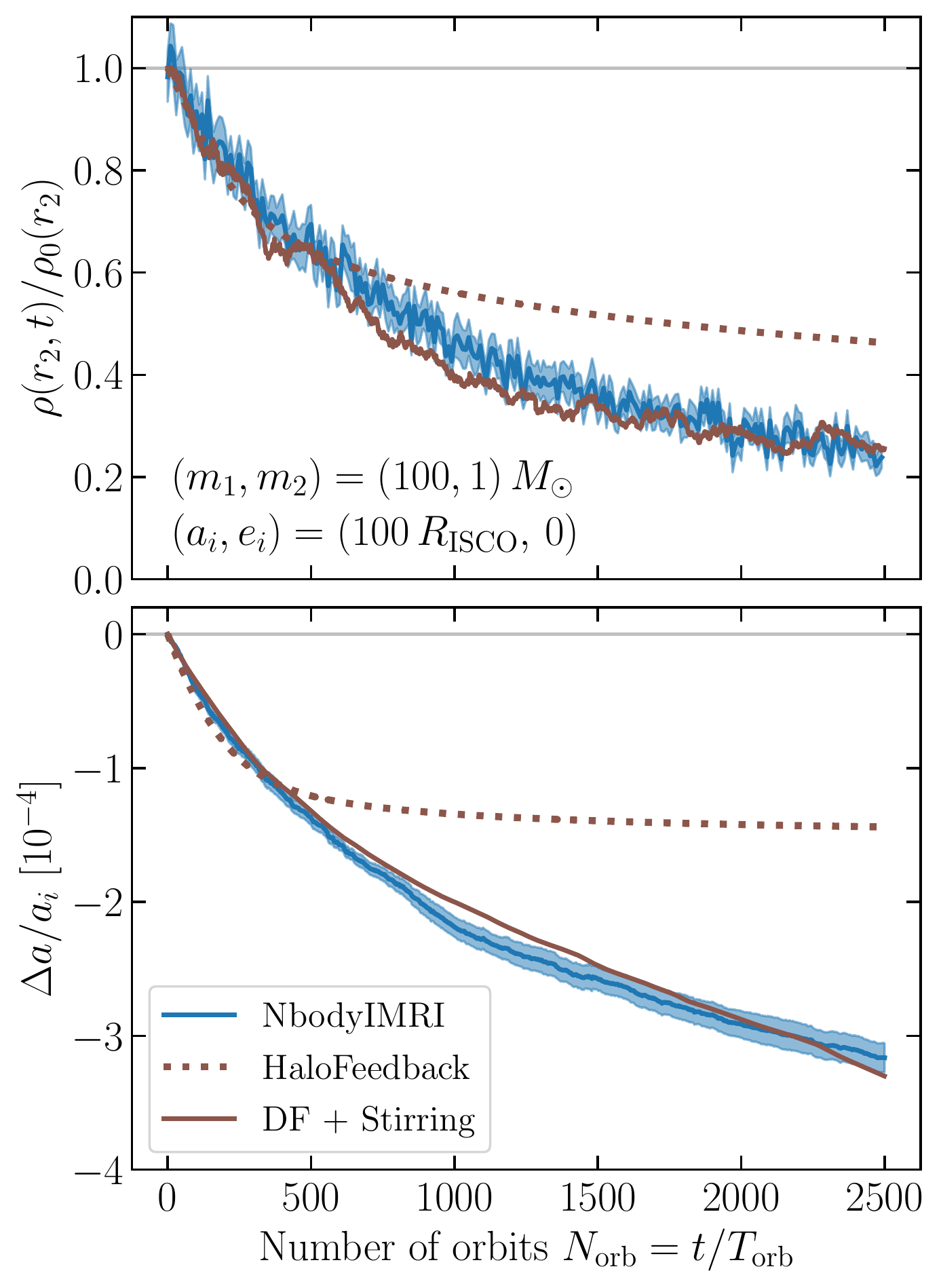}
        \caption{\textbf{Co-evolution of the DM density and semi-major axis over many orbits.} 
        In the upper panel, we plot the ratio of the DM density to the initial density at the orbital radius (vertical dashed lines in \cref{fig:Feedback}) over 2500 orbits of the binary. Simulation results are shown in blue. For comparison, we plot the prediction from \texttt{HaloFeedback} (dotted brown). A good fit to the simulation results is obtained in our toy model including feedback from both dynamical friction feedback and stirring, as described in the text (``DF + Stirring", solid brown). 
        In the lower panel, we plot the relative change in semi-major axis. The solid brown line shows the change due to dynamical friction assuming the value of $\mathcal{C}_\mathrm{DF}$ estimated in \cref{fig:DFcoefficient} and the density obtained in our ``DF + Stirring" model. 
        %While \texttt{HaloFeedback} underestimates the orbital decay, the ``DF + Stirring" model provides a good fit to the simulation, suggesting that both dynamical friction and stirring play a key role in the dynamics.
        }
    \label{fig:Feedback_time}
    \end{center}
\end{figure}

The resulting density profile is shown in the lower panel of \cref{fig:Feedback} as a solid brown line. It matches closely the DM density at the orbital radius, as well as reproducing the smooth trough in the depleted density profile out to $r = 10^3\,\risco$, increasing at larger radii. In the top panel of \cref{fig:Feedback_time}, we show the time evolution of the DM density \textit{at the orbital radius}. We see clearly that the the \texttt{HaloFeedback} formalism initially follows the results of our simulation, but quickly flattens. Instead, we see that the inclusion of the stirring mechanism (solid brown line) follows much more closely the depletion seen in our simulations, with a final density after 2500 orbits which is roughly a factor of 2 smaller than that predicted by \texttt{HaloFeedback}.

The typical amount of energy injected by close two-body interactions can be estimated as:
\begin{align}
    \langle \Delta \mathcal{E}\rangle_b &= \int P(b) \Delta \mathcal{E}(b)\,\mathrm{d}b\\
    &= 4 u^2 \left(\frac{b_{90}^2}{b_\mathrm{max}^2 - b_\mathrm{min}^2}\right) \log\Lambda\\
    &\approx 3\times 10^{-4}\,u^2\,.
\end{align}
Here, $P(b) \propto b$ is the distribution of impact parameters, and the numerical factor in the final row is valid for the specific system $(m_1,\, m_2) = (100,\,1)\,M_\odot$, with $b_\mathrm{min} \ll b_\mathrm{max}$. Noting that the characteristic energy of particles close to the secondary is $\mathcal{E} \sim \Psi(r_2) \sim u^2$, we see that the energy injected by close encounters is typically much less than this, and is therefore not sufficient to push particles to much larger radii. However, the `stirring' effect scales as $\Delta \mathcal{E} \sim r^{-5}$ and therefore injects energy preferentially into particles just outside the binary, where the energy injection $\Delta \mathcal{E}/\mathcal{E}$ can be $\mathcal{O}(1)$. The excess of particles which are pushed (by two-body interactions) to radii slightly larger than the binary separation are then pushed out further, leading to a more smoothly-varying density profile. This effect is comparable to orbital resonances in planetary systems (see e.g.~\cite{1983Natur.301..201D,1983CeMec..30..197H,1997A&A...319..290W}); orbits which would be stable in the potential of the primary BH are not stable when including the time-varying potential of the secondary object.

In the lower panel of \cref{fig:Feedback_time}, we show the evolution of the semi-major axis of the binary. The rate of decrease of the semi-major axis slows as the density profile is depleted, though the decay in $a$ continues over the 2500 orbits of the simulation. In contrast, \texttt{HaloFeedback} predicts that the decay of the semi-major axis should stall after around 500 orbits. In the \texttt{HaloFeedback} formalism, the slow-moving DM particles are rapidly depleted; we have checked explicitly that in that calculation the fraction of particles moving slower than the orbital velocity $u$ drops to about 1\% after 1000 orbits. Indeed, from the upper panel of \cref{fig:Feedback_time} it can be seen that the brown dotted line drops to roughly $1-\xi_{<u}$, because the slow-moving particles (making up a fraction $\xi_{<u} \approx 0.58$ initially) are largely depleted. This in turn would mean that the dynamical friction force, which we assume to be active only for slow-moving particles, rapidly drops to zero. 

In the lower panel of \cref{fig:Feedback_time}, we also show the decay of the semi-major axis expected in our toy model including the `stirring' effect. In this case, we calculate the dynamical friction force using the coefficient $\mathcal{C}_\mathrm{DF}$ calculated in \cref{fig:DFcoefficient} and assume a DM density following the brown solid line from the upper panel of \cref{fig:Feedback_time}. In this case, the orbital decay follows very closely the behaviour of our $N$-body simulations. This suggests that while the \textit{total} DM density in our toy model drops more rapidly than predicted by  \texttt{HaloFeedback}, the fraction of slow-moving particles must not decrease as quickly. Examining the distribution of particles in our $N$-body simulation, we indeed find that the fraction particles with $v < u$ remains approximately constant, behaviour which is matched in our toy model. Thus, the `stirring' effect not only depletes the DM spike at larger radii but must also play a role in replenishing low velocity particles on orbits close to the companion.

%---------------------------------------------
%---------------------------------------------
\section{Discussion and Conclusions}
\label{sec:Conclusions}

In this work, we have presented the publicly available code \NbodyIMRI{}~\cite{NbodyIMRI}, specifically designed for simulating binary systems embedded in cold Dark Matter spikes. The resulting simulations, covering tens to thousands of orbits, allow us to study in detail the effect of the DM spike on the evolution of the binary and the feedback of the DM in response. In a companion paper~\cite{BetterSpikesI}, we also use these simulations to study the effects of accretion on the evolution of BH binaries in DM spikes. 

We have verified that the standard dynamical friction (DF) formalism, derived assuming an infinite uniform medium, is also valid in the non-uniform systems we are considering. However, we find that the standard Chandrasekhar expression for DF, given by \cref{eq:DF_coefficient_simple}, may over-estimate the true force by 10-20\%, depending on the underlying distribution function of the DM spike. Instead, we find that the full calculation over all relative encounter velocities, given by \cref{eq:DF_coefficient}, provides a good fit to our simulation results.

We have also estimated the value of the maximum impact parameter $b_\mathrm{max}$, which ultimately determines the strength of the DF force, as well as the efficiency of feedback in the DM spike. We find that across two orders of magnitude in the mass ratio, $b_\mathrm{max} \approx (0.31 \pm 0.04)\,r_2 $ where $r_2$ is the separation of the binary.

Previous $N$-body simulations~\cite{Kavanagh:2020cfn} pointed towards a value of $b_\mathrm{max} = \sqrt{m_2/m_1}\,r_2$. For the smallest mass ratio we consider here, this would give $b_\mathrm{max} \approx 0.1\,r_2$, which is a little smaller than the value obtained in \NbodyIMRI{} simulations. For larger mass ratios, the error bars from earlier simulations made it difficult to distinguish different values of $b_\mathrm{max}$. The tailored simulations presented here have allowed us to use a factor of 8 more particles, reducing noise in the motion of the BHs. We have also been able to use shorter timesteps and therefore smaller softening lengths, in order to resolve the dynamical friction effect for larger mass ratios. Crucial to pinning down the value of $b_\mathrm{max}$ was in studying simulations with different softening lengths; $b_\mathrm{max}$ not only controls the overall normalisation of the DF force, but also the cut-off in the DF force as the softening length is increased. In \cref{sec:eccentric}, we showed further confirmation of the strength of the DF force, in simulations of binaries with eccentricities up to $e = 0.9$. 

We have also studied the feedback of the DM spike in response to circular binaries. We find that while the previous semi-analytic prescription (the \texttt{HaloFeedback} formalism presented in Ref.~\cite{Kavanagh:2020cfn}) provides a good order-of-magnitude estimate for the rate of depletion, our simulations show notable differences, including depletion out to larger radii (see \cref{fig:Feedback}). We have argued that these differences arise due to a novel `stirring' effect which arises from the time-varying potential of binary. 
This `stirring' effect does not accelerate the inspiral of the binary but has a substantial impact on the DM distribution, including replenishing slow-moving particles, which predominantly take part in dynamical friction. Including feedback from both dynamical friction and `stirring' provides a good description of our simulation results, as shown in \cref{fig:Feedback_time}.

%In order to see this feedback effect, we have reduced the number of particles in these simulations ($N_\mathrm{DM} = 16\,\mathrm{k}$) so that we can follow the system over thousands of orbits. 
While this paper was in preparation, Mukherjee \textit{et al.}~\cite{Mukherjee:2023lzn} presented results from similar $N$-body simulations. These simulations also neglect the self-gravity of the DM spike but include Post-Newtonian effects up to order 2.5 allowing them to account for both the effects of GW emission and the influence of the DM spike in their simulations. A crucial difference between \NbodyIMRI{} and the simulations of Mukherjee \textit{et al.}\ is that the latter use an adaptive time-stepping scheme, allowing them to perform simulations over much longer time-scales than presented in this paper. While our scheme uses a fixed global time step, we have chosen the size of the time step to ensure that no spuriously large two-body effects occur between the DM particles and the secondary (see \cref{app:RequiredTimestep}). The system we study in \cref{fig:Feedback_time} is comparable to the one studied in Appendix A of Mukherjee \textit{et al.}~\cite{Mukherjee:2023lzn} and we find similar behaviour, further suggesting consistency between the two sets of simulations. 

Mukherjee \textit{et al.}\ find a DM dephasing which is larger than that predicted by the standard dynamical friction formalism (including feedback) by a factor of 1-100 times. Here, we find that dynamical friction does indeed provide a good description for the energy loss in the system, as long as the novel `stirring' effect is accounted for. The lower panel of \cref{fig:Feedback_time}, for example, shows that the orbital decay due to DM is larger than expected from the \texttt{HaloFeedback} model alone, but well-described by our ``DF + Stirring" toy model. Mukherjee \textit{et al.}\ also claim to find a rate of depletion of the DM spike which is slower than that predicted by \texttt{HaloFeedback} (or rather that the fraction of slow-moving particles, which dominate the DF force, is depleted more slowly than in \texttt{HaloFeedback}). This is in fact what we have argued in \cref{sec:Feedback}: that the `stirring' mechanism replenishes slow-moving particles, such that they deplete more slowly and therefore dynamical friction remains effective for longer. 

The `stirring' we describe here is a three-body effect (arising due to the combined potential of the two compact objects acting on the DM particles, as opposed to the close two-body encounters which drive DF). Mukherjee \textit{et al.}\ similarly highlight the importance of three-body interactions, especially for DM particles close to the binary. Whether our description of the `stirring' effect is consistent with the effects observed by Mukherjee \textit{et al.}\ over long timescales is not yet clear. The much longer duration of those simulations may highlight processes which come to dominate over DF on long timescales or which may be more relevant for eccentric orbits. However, our work here highlights that both DF and `stirring' are important in understanding the co-evolution of the binary and spike. 

The \NbodyIMRI{} code is publicly available and will continue to be developed. We foresee a number of further improvements which can be made to the code. Implementing adaptive timesteps (e.g.~see \cite{1995ApJ...443L..93H,Mukherjee:2023lzn} for the specific case of a leapfrog integrator) while maintaining the timestep constraint described in \cref{app:RequiredTimestep}, would allow for a substantial speed-up of the code.  Such a speed-up would allow for these simulations to follow a much larger number of orbits and allow a more direct comparison with the results of Ref.~\cite{Mukherjee:2023lzn}.

The \NbodyIMRI{} code has been crucial in validating the semi-analytic prescriptions for compact object binaries evolving in a DM spike, as well as the subsequent feedback on the spikes themselves. These simulation results also point towards a need to extend these prescriptions, especially to include the novel `stirring' effect we have described here.

%%%%%%%%%%%%%%%%%%%%%%%%%%%%%%%%%%%
\begin{acknowledgments}
BJK acknowledges funding from the Ram\'on y Cajal Grant RYC2021-034757-I, financed by MCIN/AEI/10.13039/501100011033 and by the European Union ``NextGenerationEU"/PRTR.

TK acknowledges the support of the University of Amsterdam through the APAS-2020-GRAPPA MSc scholarship, and of Sapienza Università di Roma for support at the ``EuCAPT Workshop: Gravitational Wave probes for black hole environments".

GB gratefully acknowledges the support of the Italian Academy of Advanced Studies in America and the Department of Physics of Columbia University, where this work was finalised.

PFDC acknowledges funding from the ‘‘Fondazione Cassa di Risparmio di Firenze" under the project {\it HIPERCRHEL} for the use of high performance computing resources at the University of Florence where some of the test simulations were performed.

MP acknowledges financial support from the European Union's Horizon 2020 research and innovation program under the Marie Sk\l{}odowska-Curie grant agreement No.\ 896248.

We would like to thank Diptajyoti Mukherjee and collaborators for helpful discussions comparing our work with the results of Ref.~\cite{Mukherjee:2023lzn}.
We also acknowledge Santander Supercomputing support group at the University of Cantabria who provided access to the supercomputer Altamira Supercomputer at the Institute of Physics of Cantabria (IFCA-CSIC), member of the Spanish Supercomputing Network, for performing simulations/analyses.

\end{acknowledgments}

%%%%%%%%%%%%%%%%%%%%%%%%%%%%%%%%%%%%

\appendix

%---------------------------------------------
%---------------------------------------------
\section{Required timestep}
\label{app:RequiredTimestep}
In a softened Newtonian (Coulomb) potential, when a test particle crosses the softening radius of a mass $m$, its velocity is
\begin{equation}
    v(r = \epsilon) = \sqrt{v_\infty{}^2 + \frac{2Gm}{\epsilon}}\,,
\end{equation}
where $v_\infty$ is the initial velocity of the particle at infinity. The \NbodyIMRI{} code allows for several options for softening the potential. However, the main results we show use an `empty shell' softening approach for the force due to $m_2$: we assume that the gravitational force drops to zero for $r < \epsilon$, as if all of the compact object was concentrated on a thin shell of radius $\epsilon$. Explicitly, for a test DM particle $\tilde{m}_{\rm DM}$ with a separation $\bm{r}$ from the secondary compact object $m_2$, the force on $\tilde{m}_{\rm DM}$ is given by:
\begin{equation}
    \bm{F}_\mathrm{grav} = 
    \begin{dcases}
    - \frac{G m_2 \tilde{m}_{\rm DM}}{r^3} \bm{r} & \text{ for } r > \epsilon\,,\\
    0 &\text{ for } r \leq \epsilon\,.
    \end{dcases}
\end{equation}
With this choice of softening, the maximum velocity attained by a test particle will be the velocity at $r = \epsilon$. 

We now focus on the specific scenario of a DM pseudoparticle infalling towards the lighter compact object $m_2$ (which is itself orbiting around the heavier object $m_1$ at a separation $r_2$). The initial velocity of the infalling particle is typically of order $v_\infty \approx v_\mathrm{esc}(r_2) \approx \sqrt{2 G m_1/r_2}$. We thus have $v_\mathrm{max} \sim \sqrt{2G (m_1/r_2 + m_2/\epsilon)}$. The particle will cross a distance $\sim\epsilon$ before it leaves the `empty sphere' of the softening radius and begins decelerating, which it will do in a time:
\begin{equation}
    t_\epsilon \sim \epsilon/v_\mathrm{max} \sim \frac{\epsilon}{\sqrt{2G (m_1/r_2 + m_2/\epsilon)}}\,.
\end{equation} 
The advantage of the `empty shell' softening is that this timescale is now fixed by the softening length and the binary configuration and does not depend on the specifics of the DM particle trajectory. This places a lower bound on the dynamical timescale in this softened potential. In order to resolve this timescale with sufficient accuracy, we should have timesteps shorter than this timescale, leading to our choice of timestep:
\begin{align}
    \Delta t \leq t_\epsilon\,.
\end{align}

%---------------------------------------------
%---------------------------------------------
\section{Effects of velocity anisotropy in the spike phase-space distribution function}
\label{OManiso}

We run additional simulations using the Osipkov-Merritt models with different amounts of orbital anisotropy for the particles representing the DM spike, quantified with the usual anisotropy index (see e.g.~\cite{BinneyAndTremaine}) 
 \begin{equation}
 \label{eq:AnisotropyIndex}
\xi=\frac{2K_r}{K_t},
\end{equation}
where $K_r$ and $K_t=K_\theta+K_\phi$ are the radial and tangential components of the kinetic energy tensor.\footnote{Note that the anisotropy index $\xi$ is unrelated to the factor $\xi_{<u}$ used in the main text.} In practice, in the distribution function defined in \cref{eq:eddington}, the relative energy per unit mass $\mathcal{E}$ is substituted by  
\begin{equation}
Q=\mathcal{E}-\frac{J^2}{2r_a^2},
\end{equation}
where $J$ is the specific angular momentum of the orbit, and $\rho_{\rm DM}$ with the augmented density
\begin{equation}
\rho_{\rm DM}^*(r)=\rho_{\rm DM}(r)\left(1+\frac{r^2}{r_a^2}\right).
\end{equation}
In both expressions above $r_a$, the so-called anisotropy radius, is a control parameter quantifying the radius over which the model is dominated by low angular momentum orbits. For $\xi=1$ (corresponding to $r_a\rightarrow\infty$) the model is isotropic while the limits $\xi\rightarrow\infty$ and $\xi=0$ correspond to a model made entirely by radial or circular orbits, respectively.

\begin{figure}[tb]
    \begin{center}
        \includegraphics[width=\columnwidth]{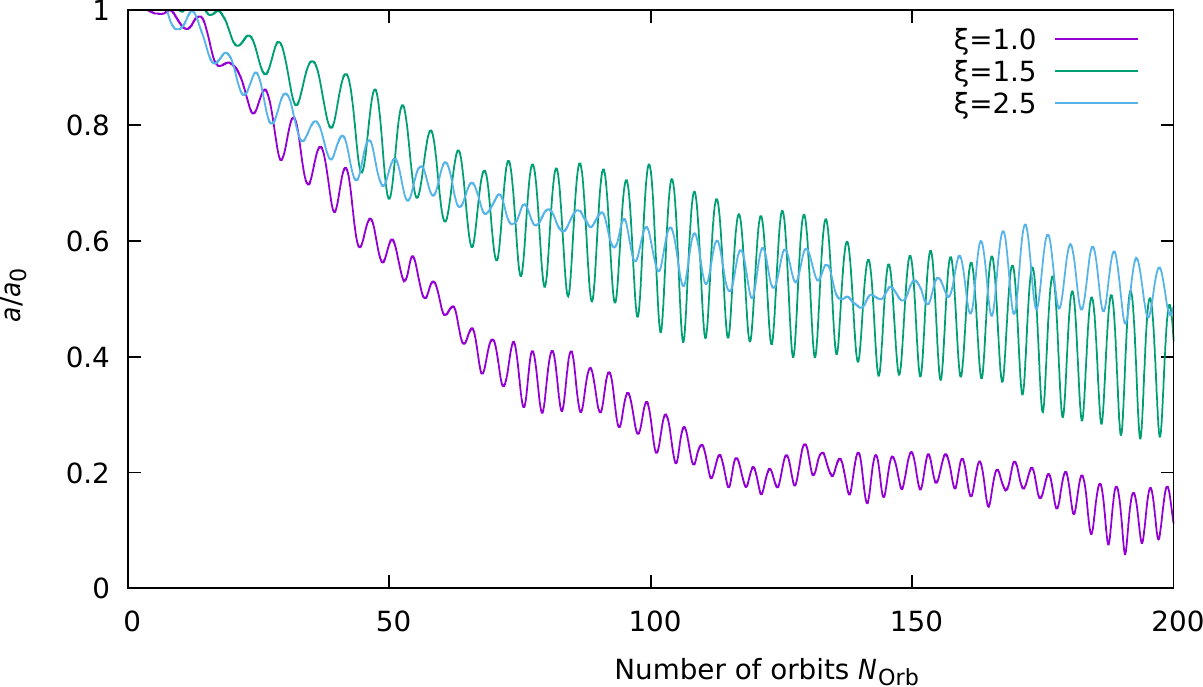}%
        \caption{\textbf{Effect of velocity anisotropy on the dynamical friction force.} Semi-major axis of the binary in units its initial value $a_0$ at $t=0$ as function of time for three different values of the anisotropy index $\xi$ of the DM particles, defined in \cref{eq:AnisotropyIndex}.}
    \label{fig:xi}
    \end{center}
\end{figure}

In the simulations, the positions of the $N_\mathrm{DM} = 256\mathrm{k}$ DM particles were as above sampled from \cref{eq:rhodm} and the velocities are then assigned from the anisotropic distribution function $f(Q_r)$ at radius $r$  with the usual rejection technique (see e.g. \cite{2017MNRAS.468.2222D} for the details). 

In \cref{fig:xi}, we show the time evolution of the semi-major axis of the binary (here $\gamma_\mathrm{sp}=2.33$ and $m_1/m_2=10^3$) for different choices of $\xi$. Increasingly radially anisotropic DM velocity distributions exert less efficient dynamical friction, for fixed DM density and binary orbital parameters. The unphysical purely tangential model (i.e., $\xi=0$, corresponding to a system with DM particle all on circular orbits, not shown here) stands out as the one with the most effective drag force.

We speculate that the form of the specific anisotropy profile of the DM spike could therefore be an important ingredient of the dynamics of an inspiraling object. Typically, spikes formed around primordial BHs~\cite{Lacki:2010zf}, thus dominated by low angular momentum orbits, should provide less friction with respect to others of comparable density and central mass $m_1$, where we expect the DM $f(v)$ to have `isotropized' via, for example, baryonic feedback processes. We leave a more detailed exploration of the effects of anisotropy to future work.

%---------------------------------------------
%---------------------------------------------
\section{`Stirring' of the DM Spike}
\label{app:Stirring}

Consider a test particle at position $\bm{r}$ with velocity $\mathbf{v}$. The potential felt by this particle is:
\begin{equation}
    \Psi(\bm{r}) = \frac{G m_1}{|\bm{r} - \bm{r}_1|} +  \frac{G m_2}{|\bm{r} - \bm{r}_2|}\,,
\end{equation}
where $\bm{r}_{1,2}$ are the positions of the primary and secondary black holes respectively. In a static potential, the specific energy $\mathcal{E} = \Psi(r) - \frac{1}{2}v^2$ of the test particle is conserved. However, in a time-dependent potential the specific energy varies over a time $\Delta t$:
\begin{equation}
    \Delta \mathcal{E} = \int_0^{\Delta t} \frac{\partial \psi}{\partial t}\,\mathrm{d}t\,.
\end{equation}
Here, we understand that the partial derivative is taken at a fixed position, while the integral over $t$ traces the trajectory of the particle over $t \in [0, \Delta t]$.

%True anomaly...
For simplicity, we consider a circular binary. We write the position vectors of the two BHs as:
\begin{align}
\bm{r}_1 &= \frac{m_2 }{m_1 + m_2} a \times (\cos\omega t, \sin\omega t, 0)\,,\\
\bm{r}_2 &= -\frac{m_1 }{m_1 + m_2} a \times (\cos\omega t, \sin\omega t, 0)\,,
\end{align}
where without loss of generality, we have taken the orbit to lie in the $x$-$y$ plane, aligned along the $x$-axis at $t = 0$. Here, the angular velocity is given by $\omega = 2\pi/T_\mathrm{orb}$. We can write:
\begin{equation}
    \left|\bm{r} - \mathbf{r_i}\right| = \left( r^2 + r_i^2  - 2 \bm{r}\cdot\bm{r}_i\right)^{1/2}\,,  
\end{equation}
and expand the potential in powers of $a/r$ (up to 3rd order):
\begin{align}
\begin{split}
    %\Psi(\bm{r}) &\approx \frac{G (m_1 + m_2)}{r} \left[1 + \right.\\
    %&\quad \left. \frac{1}{2}\frac{m_2}{m_1 + m_2}\left(\frac{a^2}{r^2}(3 \zeta^2 - 1) + \frac{a^3}{r^3} (5 \zeta^3 - 3 \zeta) \right) \right]\,.
    \Psi(\bm{r}) &\approx \frac{G (m_1 + m_2)}{r} \times \\
    &\mkern-18mu \left[ 1 + \frac{1}{2}\frac{m_2}{m_1}\left((3 \zeta^2 - 1)\frac{a^2}{r^2} + (5 \zeta^3 - 3 \zeta)\frac{a^3}{r^3}  \right) \right]\,.
\end{split}
\end{align}
Here, the orientation of the binary with respect to the test particle is encoded in:
\begin{align}
\begin{split}
\zeta = \hat{\bm{r}}\cdot\hat{\bm{r}}_1 &= \cos\omega t \sin\theta \cos\phi + \sin \omega t \sin\theta \sin\phi\\
&= \sin\theta \cos(\omega t - \phi)\,,
\end{split}
\end{align}
where $(\theta, \phi)$ are the polar coordinates of the position of the test particle.
At a fixed position, the potential varies with time as:
\begin{align}
\begin{split}
    \label{eq:partialPsi}
    \frac{\partial \psi}{\partial t}(\bm{r}) &= \frac{G (m_1 + m_2)}{r} \times \\
    &\quad \left[\frac{3}{2}\frac{m_2}{m_1}\left(2 \zeta\frac{a^2}{r^2}  + (5 \zeta^2 - 1) \right)\frac{a^3}{r^3}  \right] \frac{\partial \zeta}{\partial t}\,,
\end{split}
\end{align}
where $\partial \zeta/\partial t = -\omega \sin\theta\sin(\omega t - \phi)$.

We will now develop an approximate model for the change in energy of DM particles over a single binary orbit. We integrate $\partial\psi(\bm{r})/\partial t$ over the path of a DM particle during a time $\Delta t = T_\mathrm{orb}$. We will assume that the change in the DM particle position is small $v T_\mathrm{orb} \ll r$ over this time, which will be an increasingly good approximation as we increase $r$. We can now write the time-dependent DM coordinates as: 
\begin{align}
\begin{split}
\label{eq:AngularCoords}
r(t) &\approx r_0\left(1 + \cos\theta^\prime v_0 t/r_0\right)\,,\\
\theta(t) &\approx \theta_0 + \sin\theta^\prime \cos\phi^\prime v_0 t/r_0\,,\\
\phi(t) &\approx \phi_0 + (\sin \theta^\prime \sin\phi^\prime/\sin\theta)v_0 t/r_0\,,
\end{split}
\end{align}
where the subscript 0 denotes quantities at time $t = 0$. The angles $(\theta^\prime, \phi^\prime)$ specify the direction of the DM velocity with $\theta^\prime = 0$ pointing in the $r$ direction and $\phi^\prime = 0$ pointing in the $\theta$ direction, relative to the DM position. We now substitute \cref{eq:AngularCoords} in \cref{eq:partialPsi}, keeping only terms up to first order in $v_0 t/r_0$, and integrate over $t \in [0, T_\mathrm{orb}]$. As a final simplification, we average over $\phi^\prime \in [0, 2\pi]$, to obtain:
\begin{align}
\begin{split}
\label{eq:dE_stirring}
    \langle \Delta \mathcal{E} \rangle &= -\frac{G m_2}{r_0} \frac{v_0 T_\mathrm{orb}}{r_0}\frac{a^2}{r_0^2} \mathcal{A}(\hat{\bm{r}}, \hat{\mathbf{v}})\,, \quad \text{with,}\\
    \mathcal{A}(\hat{\bm{r}}, \hat{\mathbf{v}}) &= \frac{1}{4}\cos\theta^\prime \sin\theta_0 \times \Big[ 9\sin\theta_0\cos 2\phi_0 \\
    &\quad + \frac{a}{r_0}\left(40\cos^3\phi_0\sin^2\theta_0  - 24\cos\phi_0\right)\Big]\,.
\end{split}
\end{align}

This represents the change in energy of a DM particle outside the orbit of the binary, due to the changing potential generated by the binary. Ultimately, this energy is extracted from the binary. However, we can see that explicitly integrated the change in energy over, say, the azimuthal angle of the DM particles, we obtain zero:
\begin{equation}
    \int_0^{2\pi} \langle \Delta \mathcal{E} \rangle \,\mathrm{d}\phi_0 = 0\,.
\end{equation}
For a spherically symmetric DM spike, therefore, the net change in energy of the particles is zero. For a circular binary, this \textit{stirring} effect thus only re-distributes energy in the spike, with no net energy loss from the binary.

\bibliography{main.bib}

%merlin.mbs apsrev4-1.bst 2010-07-25 4.21a (PWD, AO, DPC) hacked
%Control: key (0)
%Control: author (0) dotless jnrlst
%Control: editor formatted (1) identically to author
%Control: production of article title (0) allowed
%Control: page (1) range
%Control: year (0) verbatim
%Control: production of eprint (0) enabled
\begin{thebibliography}{74}%
\makeatletter
\providecommand \@ifxundefined [1]{%
 \@ifx{#1\undefined}
}%
\providecommand \@ifnum [1]{%
 \ifnum #1\expandafter \@firstoftwo
 \else \expandafter \@secondoftwo
 \fi
}%
\providecommand \@ifx [1]{%
 \ifx #1\expandafter \@firstoftwo
 \else \expandafter \@secondoftwo
 \fi
}%
\providecommand \natexlab [1]{#1}%
\providecommand \enquote  [1]{``#1''}%
\providecommand \bibnamefont  [1]{#1}%
\providecommand \bibfnamefont [1]{#1}%
\providecommand \citenamefont [1]{#1}%
\providecommand \href@noop [0]{\@secondoftwo}%
\providecommand \href [0]{\begingroup \@sanitize@url \@href}%
\providecommand \@href[1]{\@@startlink{#1}\@@href}%
\providecommand \@@href[1]{\endgroup#1\@@endlink}%
\providecommand \@sanitize@url [0]{\catcode `\\12\catcode `\$12\catcode
  `\&12\catcode `\#12\catcode `\^12\catcode `\_12\catcode `\%12\relax}%
\providecommand \@@startlink[1]{}%
\providecommand \@@endlink[0]{}%
\providecommand \url  [0]{\begingroup\@sanitize@url \@url }%
\providecommand \@url [1]{\endgroup\@href {#1}{\urlprefix }}%
\providecommand \urlprefix  [0]{URL }%
\providecommand \Eprint [0]{\href }%
\providecommand \doibase [0]{http://dx.doi.org/}%
\providecommand \selectlanguage [0]{\@gobble}%
\providecommand \bibinfo  [0]{\@secondoftwo}%
\providecommand \bibfield  [0]{\@secondoftwo}%
\providecommand \translation [1]{[#1]}%
\providecommand \BibitemOpen [0]{}%
\providecommand \bibitemStop [0]{}%
\providecommand \bibitemNoStop [0]{.\EOS\space}%
\providecommand \EOS [0]{\spacefactor3000\relax}%
\providecommand \BibitemShut  [1]{\csname bibitem#1\endcsname}%
\let\auto@bib@innerbib\@empty
%</preamble>
\bibitem [{\citenamefont {Bertone}\ and\ \citenamefont
  {Tait}(2018)}]{Bertone:2018krk}%
  \BibitemOpen
  \bibfield  {author} {\bibinfo {author} {\bibfnamefont {Gianfranco}\
  \bibnamefont {Bertone}}\ and\ \bibinfo {author} {\bibfnamefont {Tim}\
  \bibnamefont {Tait}, \bibfnamefont {M.~P.}},\ }\bibfield  {title} {\enquote
  {\bibinfo {title} {{A new era in the search for dark matter}},}\ }\href
  {\doibase 10.1038/s41586-018-0542-z} {\bibfield  {journal} {\bibinfo
  {journal} {Nature}\ }\textbf {\bibinfo {volume} {562}},\ \bibinfo {pages}
  {51--56} (\bibinfo {year} {2018})},\ \Eprint
  {http://arxiv.org/abs/1810.01668} {arXiv:1810.01668 [astro-ph.CO]}
  \BibitemShut {NoStop}%
\bibitem [{\citenamefont {Bertone}\ \emph {et~al.}(2020)\citenamefont {Bertone}
  \emph {et~al.}}]{Bertone:2019irm}%
  \BibitemOpen
  \bibfield  {author} {\bibinfo {author} {\bibfnamefont {Gianfranco}\
  \bibnamefont {Bertone}} \emph {et~al.},\ }\bibfield  {title} {\enquote
  {\bibinfo {title} {{Gravitational wave probes of dark matter: challenges and
  opportunities}},}\ }\href {\doibase 10.21468/SciPostPhysCore.3.2.007}
  {\bibfield  {journal} {\bibinfo  {journal} {SciPost Phys. Core}\ }\textbf
  {\bibinfo {volume} {3}},\ \bibinfo {pages} {007} (\bibinfo {year} {2020})},\
  \Eprint {http://arxiv.org/abs/1907.10610} {arXiv:1907.10610 [astro-ph.CO]}
  \BibitemShut {NoStop}%
\bibitem [{\citenamefont {Baryakhtar}\ \emph {et~al.}(2022)\citenamefont
  {Baryakhtar} \emph {et~al.}}]{Baryakhtar:2022hbu}%
  \BibitemOpen
  \bibfield  {author} {\bibinfo {author} {\bibfnamefont {Masha}\ \bibnamefont
  {Baryakhtar}} \emph {et~al.},\ }\bibfield  {title} {\enquote {\bibinfo
  {title} {{Dark Matter In Extreme Astrophysical Environments}},}\ }in\
  \href@noop {} {\emph {\bibinfo {booktitle} {{Snowmass 2021}}}}\ (\bibinfo
  {year} {2022})\ \Eprint {http://arxiv.org/abs/2203.07984} {arXiv:2203.07984
  [hep-ph]} \BibitemShut {NoStop}%
\bibitem [{\citenamefont {Amaro-Seoane}\ \emph {et~al.}(2017)\citenamefont
  {Amaro-Seoane} \emph {et~al.}}]{LISA}%
  \BibitemOpen
  \bibfield  {author} {\bibinfo {author} {\bibfnamefont {Pau}\ \bibnamefont
  {Amaro-Seoane}} \emph {et~al.} (\bibinfo {collaboration} {LISA}),\
  }\href@noop {} {\enquote {\bibinfo {title} {{Laser Interferometer Space
  Antenna}},}\ } (\bibinfo {year} {2017}),\ \Eprint
  {http://arxiv.org/abs/1702.00786} {arXiv:1702.00786 [astro-ph.IM]}
  \BibitemShut {NoStop}%
\bibitem [{\citenamefont {Hu}\ and\ \citenamefont {Wu}(2017)}]{Hu:2017mde}%
  \BibitemOpen
  \bibfield  {author} {\bibinfo {author} {\bibfnamefont {Wen-Rui}\ \bibnamefont
  {Hu}}\ and\ \bibinfo {author} {\bibfnamefont {Yue-Liang}\ \bibnamefont
  {Wu}},\ }\bibfield  {title} {\enquote {\bibinfo {title} {{The Taiji Program
  in Space for gravitational wave physics and the nature of gravity}},}\ }\href
  {\doibase 10.1093/nsr/nwx116} {\bibfield  {journal} {\bibinfo  {journal}
  {Natl. Sci. Rev.}\ }\textbf {\bibinfo {volume} {4}},\ \bibinfo {pages}
  {685--686} (\bibinfo {year} {2017})}\BibitemShut {NoStop}%
\bibitem [{\citenamefont {Gondolo}\ and\ \citenamefont
  {Silk}(1999)}]{Gondolo:1999ef}%
  \BibitemOpen
  \bibfield  {author} {\bibinfo {author} {\bibfnamefont {Paolo}\ \bibnamefont
  {Gondolo}}\ and\ \bibinfo {author} {\bibfnamefont {Joseph}\ \bibnamefont
  {Silk}},\ }\bibfield  {title} {\enquote {\bibinfo {title} {{Dark matter
  annihilation at the galactic center}},}\ }\href {\doibase
  10.1103/PhysRevLett.83.1719} {\bibfield  {journal} {\bibinfo  {journal}
  {Phys. Rev. Lett.}\ }\textbf {\bibinfo {volume} {83}},\ \bibinfo {pages}
  {1719--1722} (\bibinfo {year} {1999})},\ \Eprint
  {http://arxiv.org/abs/astro-ph/9906391} {arXiv:astro-ph/9906391} \BibitemShut
  {NoStop}%
\bibitem [{\citenamefont {Ullio}\ \emph {et~al.}(2001)\citenamefont {Ullio},
  \citenamefont {Zhao},\ and\ \citenamefont {Kamionkowski}}]{Ullio:2001fb}%
  \BibitemOpen
  \bibfield  {author} {\bibinfo {author} {\bibfnamefont {Piero}\ \bibnamefont
  {Ullio}}, \bibinfo {author} {\bibfnamefont {HongSheng}\ \bibnamefont {Zhao}},
  \ and\ \bibinfo {author} {\bibfnamefont {Marc}\ \bibnamefont
  {Kamionkowski}},\ }\bibfield  {title} {\enquote {\bibinfo {title} {{A Dark
  matter spike at the galactic center?}}}\ }\href {\doibase
  10.1103/PhysRevD.64.043504} {\bibfield  {journal} {\bibinfo  {journal} {Phys.
  Rev. D}\ }\textbf {\bibinfo {volume} {64}},\ \bibinfo {pages} {043504}
  (\bibinfo {year} {2001})},\ \Eprint {http://arxiv.org/abs/astro-ph/0101481}
  {arXiv:astro-ph/0101481} \BibitemShut {NoStop}%
\bibitem [{\citenamefont {Bertone}\ and\ \citenamefont
  {Merritt}(2005)}]{Bertone:2005hw}%
  \BibitemOpen
  \bibfield  {author} {\bibinfo {author} {\bibfnamefont {Gianfranco}\
  \bibnamefont {Bertone}}\ and\ \bibinfo {author} {\bibfnamefont {David}\
  \bibnamefont {Merritt}},\ }\bibfield  {title} {\enquote {\bibinfo {title}
  {{Time-dependent models for dark matter at the Galactic Center}},}\ }\href
  {\doibase 10.1103/PhysRevD.72.103502} {\bibfield  {journal} {\bibinfo
  {journal} {Phys. Rev. D}\ }\textbf {\bibinfo {volume} {72}},\ \bibinfo
  {pages} {103502} (\bibinfo {year} {2005})},\ \Eprint
  {http://arxiv.org/abs/astro-ph/0501555} {arXiv:astro-ph/0501555} \BibitemShut
  {NoStop}%
\bibitem [{\citenamefont {Mack}\ \emph {et~al.}(2007)\citenamefont {Mack},
  \citenamefont {Ostriker},\ and\ \citenamefont {Ricotti}}]{Mack:2006gz}%
  \BibitemOpen
  \bibfield  {author} {\bibinfo {author} {\bibfnamefont {Katherine~J.}\
  \bibnamefont {Mack}}, \bibinfo {author} {\bibfnamefont {Jeremiah~P.}\
  \bibnamefont {Ostriker}}, \ and\ \bibinfo {author} {\bibfnamefont {Massimo}\
  \bibnamefont {Ricotti}},\ }\bibfield  {title} {\enquote {\bibinfo {title}
  {{Growth of structure seeded by primordial black holes}},}\ }\href {\doibase
  10.1086/518998} {\bibfield  {journal} {\bibinfo  {journal} {Astrophys. J.}\
  }\textbf {\bibinfo {volume} {665}},\ \bibinfo {pages} {1277--1287} (\bibinfo
  {year} {2007})},\ \Eprint {http://arxiv.org/abs/astro-ph/0608642}
  {arXiv:astro-ph/0608642} \BibitemShut {NoStop}%
\bibitem [{\citenamefont {Eroshenko}(2016)}]{Eroshenko:2016yve}%
  \BibitemOpen
  \bibfield  {author} {\bibinfo {author} {\bibfnamefont {Yu.~N.}\ \bibnamefont
  {Eroshenko}},\ }\bibfield  {title} {\enquote {\bibinfo {title} {{Dark matter
  density spikes around primordial black holes}},}\ }\href {\doibase
  10.1134/S1063773716060013} {\bibfield  {journal} {\bibinfo  {journal}
  {Astron. Lett.}\ }\textbf {\bibinfo {volume} {42}},\ \bibinfo {pages}
  {347--356} (\bibinfo {year} {2016})},\ \Eprint
  {http://arxiv.org/abs/1607.00612} {arXiv:1607.00612 [astro-ph.HE]}
  \BibitemShut {NoStop}%
\bibitem [{\citenamefont {Adamek}\ \emph {et~al.}(2019)\citenamefont {Adamek},
  \citenamefont {Byrnes}, \citenamefont {Gosenca},\ and\ \citenamefont
  {Hotchkiss}}]{Adamek:2019gns}%
  \BibitemOpen
  \bibfield  {author} {\bibinfo {author} {\bibfnamefont {Julian}\ \bibnamefont
  {Adamek}}, \bibinfo {author} {\bibfnamefont {Christian~T.}\ \bibnamefont
  {Byrnes}}, \bibinfo {author} {\bibfnamefont {Mateja}\ \bibnamefont
  {Gosenca}}, \ and\ \bibinfo {author} {\bibfnamefont {Shaun}\ \bibnamefont
  {Hotchkiss}},\ }\bibfield  {title} {\enquote {\bibinfo {title} {{WIMPs and
  stellar-mass primordial black holes are incompatible}},}\ }\href {\doibase
  10.1103/PhysRevD.100.023506} {\bibfield  {journal} {\bibinfo  {journal}
  {Phys. Rev. D}\ }\textbf {\bibinfo {volume} {100}},\ \bibinfo {pages}
  {023506} (\bibinfo {year} {2019})},\ \Eprint
  {http://arxiv.org/abs/1901.08528} {arXiv:1901.08528 [astro-ph.CO]}
  \BibitemShut {NoStop}%
\bibitem [{\citenamefont {Green}\ and\ \citenamefont
  {Kavanagh}(2021)}]{Green:2020jor}%
  \BibitemOpen
  \bibfield  {author} {\bibinfo {author} {\bibfnamefont {Anne~M.}\ \bibnamefont
  {Green}}\ and\ \bibinfo {author} {\bibfnamefont {Bradley~J.}\ \bibnamefont
  {Kavanagh}},\ }\bibfield  {title} {\enquote {\bibinfo {title} {{Primordial
  Black Holes as a dark matter candidate}},}\ }\href {\doibase
  10.1088/1361-6471/abc534} {\bibfield  {journal} {\bibinfo  {journal} {J.
  Phys. G}\ }\textbf {\bibinfo {volume} {48}},\ \bibinfo {pages} {043001}
  (\bibinfo {year} {2021})},\ \Eprint {http://arxiv.org/abs/2007.10722}
  {arXiv:2007.10722 [astro-ph.CO]} \BibitemShut {NoStop}%
\bibitem [{\citenamefont {{Eda}}\ \emph {et~al.}(2013)\citenamefont {{Eda}},
  \citenamefont {{Itoh}}, \citenamefont {{Kuroyanagi}},\ and\ \citenamefont
  {{Silk}}}]{Eda1}%
  \BibitemOpen
  \bibfield  {author} {\bibinfo {author} {\bibfnamefont {Kazunari}\
  \bibnamefont {{Eda}}}, \bibinfo {author} {\bibfnamefont {Yousuke}\
  \bibnamefont {{Itoh}}}, \bibinfo {author} {\bibfnamefont {Sachiko}\
  \bibnamefont {{Kuroyanagi}}}, \ and\ \bibinfo {author} {\bibfnamefont
  {Joseph}\ \bibnamefont {{Silk}}},\ }\bibfield  {title} {\enquote {\bibinfo
  {title} {{New Probe of Dark-Matter Properties: Gravitational Waves from an
  Intermediate-Mass Black Hole Embedded in a Dark-Matter Minispike}},}\ }\href
  {\doibase 10.1103/PhysRevLett.110.221101} {\bibfield  {journal} {\bibinfo
  {journal} {\prl}\ }\textbf {\bibinfo {volume} {110}},\ \bibinfo {eid}
  {221101} (\bibinfo {year} {2013})},\ \Eprint {http://arxiv.org/abs/1301.5971}
  {arXiv:1301.5971 [gr-qc]} \BibitemShut {NoStop}%
\bibitem [{\citenamefont {{Eda}}\ \emph {et~al.}(2015)\citenamefont {{Eda}},
  \citenamefont {{Itoh}}, \citenamefont {{Kuroyanagi}},\ and\ \citenamefont
  {{Silk}}}]{Eda2}%
  \BibitemOpen
  \bibfield  {author} {\bibinfo {author} {\bibfnamefont {Kazunari}\
  \bibnamefont {{Eda}}}, \bibinfo {author} {\bibfnamefont {Yousuke}\
  \bibnamefont {{Itoh}}}, \bibinfo {author} {\bibfnamefont {Sachiko}\
  \bibnamefont {{Kuroyanagi}}}, \ and\ \bibinfo {author} {\bibfnamefont
  {Joseph}\ \bibnamefont {{Silk}}},\ }\bibfield  {title} {\enquote {\bibinfo
  {title} {{Gravitational waves as a probe of dark matter minispikes}},}\
  }\href {\doibase 10.1103/PhysRevD.91.044045} {\bibfield  {journal} {\bibinfo
  {journal} {\prd}\ }\textbf {\bibinfo {volume} {91}},\ \bibinfo {eid} {044045}
  (\bibinfo {year} {2015})},\ \Eprint {http://arxiv.org/abs/1408.3534}
  {arXiv:1408.3534 [gr-qc]} \BibitemShut {NoStop}%
\bibitem [{\citenamefont {Macedo}\ \emph {et~al.}(2013)\citenamefont {Macedo},
  \citenamefont {Pani}, \citenamefont {Cardoso},\ and\ \citenamefont
  {Crispino}}]{Macedo:2013qea}%
  \BibitemOpen
  \bibfield  {author} {\bibinfo {author} {\bibfnamefont {Caio F.~B.}\
  \bibnamefont {Macedo}}, \bibinfo {author} {\bibfnamefont {Paolo}\
  \bibnamefont {Pani}}, \bibinfo {author} {\bibfnamefont {Vitor}\ \bibnamefont
  {Cardoso}}, \ and\ \bibinfo {author} {\bibfnamefont {Lu\'\i{}s C.~B.}\
  \bibnamefont {Crispino}},\ }\bibfield  {title} {\enquote {\bibinfo {title}
  {{Into the lair: gravitational-wave signatures of dark matter}},}\ }\href
  {\doibase 10.1088/0004-637X/774/1/48} {\bibfield  {journal} {\bibinfo
  {journal} {Astrophys. J.}\ }\textbf {\bibinfo {volume} {774}},\ \bibinfo
  {pages} {48} (\bibinfo {year} {2013})},\ \Eprint
  {http://arxiv.org/abs/1302.2646} {arXiv:1302.2646 [gr-qc]} \BibitemShut
  {NoStop}%
\bibitem [{\citenamefont {Barausse}\ \emph {et~al.}(2014)\citenamefont
  {Barausse}, \citenamefont {Cardoso},\ and\ \citenamefont
  {Pani}}]{Barausse:2014tra}%
  \BibitemOpen
  \bibfield  {author} {\bibinfo {author} {\bibfnamefont {Enrico}\ \bibnamefont
  {Barausse}}, \bibinfo {author} {\bibfnamefont {Vitor}\ \bibnamefont
  {Cardoso}}, \ and\ \bibinfo {author} {\bibfnamefont {Paolo}\ \bibnamefont
  {Pani}},\ }\bibfield  {title} {\enquote {\bibinfo {title} {{Can environmental
  effects spoil precision gravitational-wave astrophysics?}}}\ }\href {\doibase
  10.1103/PhysRevD.89.104059} {\bibfield  {journal} {\bibinfo  {journal} {Phys.
  Rev. D}\ }\textbf {\bibinfo {volume} {89}},\ \bibinfo {pages} {104059}
  (\bibinfo {year} {2014})},\ \Eprint {http://arxiv.org/abs/1404.7149}
  {arXiv:1404.7149 [gr-qc]} \BibitemShut {NoStop}%
\bibitem [{\citenamefont {Barausse}\ \emph {et~al.}(2015)\citenamefont
  {Barausse}, \citenamefont {Cardoso},\ and\ \citenamefont
  {Pani}}]{Barausse:2014pra}%
  \BibitemOpen
  \bibfield  {author} {\bibinfo {author} {\bibfnamefont {Enrico}\ \bibnamefont
  {Barausse}}, \bibinfo {author} {\bibfnamefont {Vitor}\ \bibnamefont
  {Cardoso}}, \ and\ \bibinfo {author} {\bibfnamefont {Paolo}\ \bibnamefont
  {Pani}},\ }\bibfield  {title} {\enquote {\bibinfo {title} {{Environmental
  Effects for Gravitational-wave Astrophysics}},}\ }\href {\doibase
  10.1088/1742-6596/610/1/012044} {\bibfield  {journal} {\bibinfo  {journal}
  {J. Phys. Conf. Ser.}\ }\textbf {\bibinfo {volume} {610}},\ \bibinfo {pages}
  {012044} (\bibinfo {year} {2015})},\ \Eprint {http://arxiv.org/abs/1404.7140}
  {arXiv:1404.7140 [astro-ph.CO]} \BibitemShut {NoStop}%
\bibitem [{\citenamefont {{Alexander}}\ and\ \citenamefont
  {{Pfuhl}}(2014)}]{2014ApJ...780..148A}%
  \BibitemOpen
  \bibfield  {author} {\bibinfo {author} {\bibfnamefont {Tal}\ \bibnamefont
  {{Alexander}}}\ and\ \bibinfo {author} {\bibfnamefont {Oliver}\ \bibnamefont
  {{Pfuhl}}},\ }\bibfield  {title} {\enquote {\bibinfo {title} {{Constraining
  the Dark Cusp in the Galactic Center by Long-period Binaries}},}\ }\href
  {\doibase 10.1088/0004-637X/780/2/148} {\bibfield  {journal} {\bibinfo
  {journal} {ApJ}\ }\textbf {\bibinfo {volume} {780}},\ \bibinfo {eid} {148}
  (\bibinfo {year} {2014})},\ \Eprint {http://arxiv.org/abs/1308.6638}
  {arXiv:1308.6638 [astro-ph.GA]} \BibitemShut {NoStop}%
\bibitem [{\citenamefont {Becker}\ and\ \citenamefont
  {Sagunski}(2023)}]{Becker:2022wlo}%
  \BibitemOpen
  \bibfield  {author} {\bibinfo {author} {\bibfnamefont {Niklas}\ \bibnamefont
  {Becker}}\ and\ \bibinfo {author} {\bibfnamefont {Laura}\ \bibnamefont
  {Sagunski}},\ }\bibfield  {title} {\enquote {\bibinfo {title} {{Comparing
  accretion disks and dark matter spikes in intermediate mass ratio
  inspirals}},}\ }\href {\doibase 10.1103/PhysRevD.107.083003} {\bibfield
  {journal} {\bibinfo  {journal} {Phys. Rev. D}\ }\textbf {\bibinfo {volume}
  {107}},\ \bibinfo {pages} {083003} (\bibinfo {year} {2023})},\ \Eprint
  {http://arxiv.org/abs/2211.05145} {arXiv:2211.05145 [gr-qc]} \BibitemShut
  {NoStop}%
\bibitem [{\citenamefont {Cole}\ \emph
  {et~al.}(2023{\natexlab{a}})\citenamefont {Cole}, \citenamefont {Bertone},
  \citenamefont {Coogan}, \citenamefont {Gaggero}, \citenamefont {Karydas},
  \citenamefont {Kavanagh}, \citenamefont {Spieksma},\ and\ \citenamefont
  {Tomaselli}}]{Cole:2022yzw}%
  \BibitemOpen
  \bibfield  {author} {\bibinfo {author} {\bibfnamefont {Philippa~S.}\
  \bibnamefont {Cole}}, \bibinfo {author} {\bibfnamefont {Gianfranco}\
  \bibnamefont {Bertone}}, \bibinfo {author} {\bibfnamefont {Adam}\
  \bibnamefont {Coogan}}, \bibinfo {author} {\bibfnamefont {Daniele}\
  \bibnamefont {Gaggero}}, \bibinfo {author} {\bibfnamefont {Theophanes}\
  \bibnamefont {Karydas}}, \bibinfo {author} {\bibfnamefont {Bradley~J.}\
  \bibnamefont {Kavanagh}}, \bibinfo {author} {\bibfnamefont {Thomas F.~M.}\
  \bibnamefont {Spieksma}}, \ and\ \bibinfo {author} {\bibfnamefont
  {Giovanni~Maria}\ \bibnamefont {Tomaselli}},\ }\bibfield  {title} {\enquote
  {\bibinfo {title} {{Distinguishing environmental effects on binary black hole
  gravitational waveforms}},}\ }\href {\doibase 10.1038/s41550-023-01990-2}
  {\bibfield  {journal} {\bibinfo  {journal} {Nature Astron.}\ }\textbf
  {\bibinfo {volume} {7}},\ \bibinfo {pages} {943--950} (\bibinfo {year}
  {2023}{\natexlab{a}})},\ \Eprint {http://arxiv.org/abs/2211.01362}
  {arXiv:2211.01362 [gr-qc]} \BibitemShut {NoStop}%
\bibitem [{\citenamefont {Cardoso}\ and\ \citenamefont
  {Maselli}(2020)}]{Cardoso:2019rou}%
  \BibitemOpen
  \bibfield  {author} {\bibinfo {author} {\bibfnamefont {Vitor}\ \bibnamefont
  {Cardoso}}\ and\ \bibinfo {author} {\bibfnamefont {Andrea}\ \bibnamefont
  {Maselli}},\ }\bibfield  {title} {\enquote {\bibinfo {title} {{Constraints on
  the astrophysical environment of binaries with gravitational-wave
  observations}},}\ }\href {\doibase 10.1051/0004-6361/202037654} {\bibfield
  {journal} {\bibinfo  {journal} {Astron. Astrophys.}\ }\textbf {\bibinfo
  {volume} {644}},\ \bibinfo {pages} {A147} (\bibinfo {year} {2020})},\ \Eprint
  {http://arxiv.org/abs/1909.05870} {arXiv:1909.05870 [astro-ph.HE]}
  \BibitemShut {NoStop}%
\bibitem [{\citenamefont {Edwards}\ \emph {et~al.}(2020)\citenamefont
  {Edwards}, \citenamefont {Chianese}, \citenamefont {Kavanagh}, \citenamefont
  {Nissanke},\ and\ \citenamefont {Weniger}}]{Edwards:2019tzf}%
  \BibitemOpen
  \bibfield  {author} {\bibinfo {author} {\bibfnamefont {Thomas D.~P.}\
  \bibnamefont {Edwards}}, \bibinfo {author} {\bibfnamefont {Marco}\
  \bibnamefont {Chianese}}, \bibinfo {author} {\bibfnamefont {Bradley~J.}\
  \bibnamefont {Kavanagh}}, \bibinfo {author} {\bibfnamefont {Samaya~M.}\
  \bibnamefont {Nissanke}}, \ and\ \bibinfo {author} {\bibfnamefont
  {Christoph}\ \bibnamefont {Weniger}},\ }\bibfield  {title} {\enquote
  {\bibinfo {title} {{Unique Multimessenger Signal of QCD Axion Dark
  Matter}},}\ }\href {\doibase 10.1103/PhysRevLett.124.161101} {\bibfield
  {journal} {\bibinfo  {journal} {Phys. Rev. Lett.}\ }\textbf {\bibinfo
  {volume} {124}},\ \bibinfo {pages} {161101} (\bibinfo {year} {2020})},\
  \Eprint {http://arxiv.org/abs/1905.04686} {arXiv:1905.04686 [hep-ph]}
  \BibitemShut {NoStop}%
\bibitem [{\citenamefont {Coogan}\ \emph {et~al.}(2022)\citenamefont {Coogan},
  \citenamefont {Bertone}, \citenamefont {Gaggero}, \citenamefont {Kavanagh},\
  and\ \citenamefont {Nichols}}]{Coogan:2021uqv}%
  \BibitemOpen
  \bibfield  {author} {\bibinfo {author} {\bibfnamefont {Adam}\ \bibnamefont
  {Coogan}}, \bibinfo {author} {\bibfnamefont {Gianfranco}\ \bibnamefont
  {Bertone}}, \bibinfo {author} {\bibfnamefont {Daniele}\ \bibnamefont
  {Gaggero}}, \bibinfo {author} {\bibfnamefont {Bradley~J.}\ \bibnamefont
  {Kavanagh}}, \ and\ \bibinfo {author} {\bibfnamefont {David~A.}\ \bibnamefont
  {Nichols}},\ }\bibfield  {title} {\enquote {\bibinfo {title} {{Measuring the
  dark matter environments of black hole binaries with gravitational waves}},}\
  }\href {\doibase 10.1103/PhysRevD.105.043009} {\bibfield  {journal} {\bibinfo
   {journal} {Phys. Rev. D}\ }\textbf {\bibinfo {volume} {105}},\ \bibinfo
  {pages} {043009} (\bibinfo {year} {2022})},\ \Eprint
  {http://arxiv.org/abs/2108.04154} {arXiv:2108.04154 [gr-qc]} \BibitemShut
  {NoStop}%
\bibitem [{\citenamefont {Cole}\ \emph
  {et~al.}(2023{\natexlab{b}})\citenamefont {Cole}, \citenamefont {Coogan},
  \citenamefont {Kavanagh},\ and\ \citenamefont {Bertone}}]{Cole:2022ucw}%
  \BibitemOpen
  \bibfield  {author} {\bibinfo {author} {\bibfnamefont {Philippa~S.}\
  \bibnamefont {Cole}}, \bibinfo {author} {\bibfnamefont {Adam}\ \bibnamefont
  {Coogan}}, \bibinfo {author} {\bibfnamefont {Bradley~J.}\ \bibnamefont
  {Kavanagh}}, \ and\ \bibinfo {author} {\bibfnamefont {Gianfranco}\
  \bibnamefont {Bertone}},\ }\bibfield  {title} {\enquote {\bibinfo {title}
  {{Measuring dark matter spikes around primordial black holes with Einstein
  Telescope and Cosmic Explorer}},}\ }\href {\doibase
  10.1103/PhysRevD.107.083006} {\bibfield  {journal} {\bibinfo  {journal}
  {Phys. Rev. D}\ }\textbf {\bibinfo {volume} {107}},\ \bibinfo {pages}
  {083006} (\bibinfo {year} {2023}{\natexlab{b}})},\ \Eprint
  {http://arxiv.org/abs/2207.07576} {arXiv:2207.07576 [astro-ph.CO]}
  \BibitemShut {NoStop}%
\bibitem [{\citenamefont {Zhang}\ \emph {et~al.}(2024)\citenamefont {Zhang},
  \citenamefont {Fu},\ and\ \citenamefont {Dai}}]{Zhang:2024ugv}%
  \BibitemOpen
  \bibfield  {author} {\bibinfo {author} {\bibfnamefont {Chao}\ \bibnamefont
  {Zhang}}, \bibinfo {author} {\bibfnamefont {Guoyang}\ \bibnamefont {Fu}}, \
  and\ \bibinfo {author} {\bibfnamefont {Ning}\ \bibnamefont {Dai}},\
  }\bibfield  {title} {\enquote {\bibinfo {title} {{Detecting dark matter with
  extreme mass-ratio inspirals}},}\ }\href@noop {} {\  (\bibinfo {year}
  {2024})},\ \Eprint {http://arxiv.org/abs/2401.04467} {arXiv:2401.04467
  [gr-qc]} \BibitemShut {NoStop}%
\bibitem [{\citenamefont {Yue}\ and\ \citenamefont {Han}(2018)}]{Yue:2017iwc}%
  \BibitemOpen
  \bibfield  {author} {\bibinfo {author} {\bibfnamefont {Xiao-Jun}\
  \bibnamefont {Yue}}\ and\ \bibinfo {author} {\bibfnamefont {Wen-Biao}\
  \bibnamefont {Han}},\ }\bibfield  {title} {\enquote {\bibinfo {title}
  {{Gravitational waves with dark matter minispikes: the combined effect}},}\
  }\href {\doibase 10.1103/PhysRevD.97.064003} {\bibfield  {journal} {\bibinfo
  {journal} {Phys. Rev. D}\ }\textbf {\bibinfo {volume} {97}},\ \bibinfo
  {pages} {064003} (\bibinfo {year} {2018})},\ \Eprint
  {http://arxiv.org/abs/1711.09706} {arXiv:1711.09706 [gr-qc]} \BibitemShut
  {NoStop}%
\bibitem [{\citenamefont {Yue}\ \emph {et~al.}(2019)\citenamefont {Yue},
  \citenamefont {Han},\ and\ \citenamefont {Chen}}]{Yue:2018vtk}%
  \BibitemOpen
  \bibfield  {author} {\bibinfo {author} {\bibfnamefont {Xiao-Jun}\
  \bibnamefont {Yue}}, \bibinfo {author} {\bibfnamefont {Wen-Biao}\
  \bibnamefont {Han}}, \ and\ \bibinfo {author} {\bibfnamefont {Xian}\
  \bibnamefont {Chen}},\ }\bibfield  {title} {\enquote {\bibinfo {title} {{Dark
  matter: an efficient catalyst for intermediate-mass-ratio-inspiral
  events}},}\ }\href {\doibase 10.3847/1538-4357/ab06f6} {\bibfield  {journal}
  {\bibinfo  {journal} {Astrophys. J.}\ }\textbf {\bibinfo {volume} {874}},\
  \bibinfo {pages} {34} (\bibinfo {year} {2019})},\ \Eprint
  {http://arxiv.org/abs/1802.03739} {arXiv:1802.03739 [gr-qc]} \BibitemShut
  {NoStop}%
\bibitem [{\citenamefont {Hannuksela}\ \emph {et~al.}(2020)\citenamefont
  {Hannuksela}, \citenamefont {Ng},\ and\ \citenamefont
  {Li}}]{Hannuksela:2019vip}%
  \BibitemOpen
  \bibfield  {author} {\bibinfo {author} {\bibfnamefont {Otto~A.}\ \bibnamefont
  {Hannuksela}}, \bibinfo {author} {\bibfnamefont {Kenny C.~Y.}\ \bibnamefont
  {Ng}}, \ and\ \bibinfo {author} {\bibfnamefont {Tjonnie G.~F.}\ \bibnamefont
  {Li}},\ }\bibfield  {title} {\enquote {\bibinfo {title} {{Extreme dark matter
  tests with extreme mass ratio inspirals}},}\ }\href {\doibase
  10.1103/PhysRevD.102.103022} {\bibfield  {journal} {\bibinfo  {journal}
  {Phys. Rev. D}\ }\textbf {\bibinfo {volume} {102}},\ \bibinfo {pages}
  {103022} (\bibinfo {year} {2020})},\ \Eprint
  {http://arxiv.org/abs/1906.11845} {arXiv:1906.11845 [astro-ph.CO]}
  \BibitemShut {NoStop}%
\bibitem [{\citenamefont {Nichols}\ \emph {et~al.}(2023)\citenamefont
  {Nichols}, \citenamefont {Wade},\ and\ \citenamefont
  {Grant}}]{Nichols:2023ufs}%
  \BibitemOpen
  \bibfield  {author} {\bibinfo {author} {\bibfnamefont {David~A.}\
  \bibnamefont {Nichols}}, \bibinfo {author} {\bibfnamefont {Benjamin~A.}\
  \bibnamefont {Wade}}, \ and\ \bibinfo {author} {\bibfnamefont {Alexander~M.}\
  \bibnamefont {Grant}},\ }\bibfield  {title} {\enquote {\bibinfo {title}
  {{Secondary accretion of dark matter in intermediate mass-ratio inspirals:
  Dark-matter dynamics and gravitational-wave phase}},}\ }\href {\doibase
  10.1103/PhysRevD.108.124062} {\bibfield  {journal} {\bibinfo  {journal}
  {Phys. Rev. D}\ }\textbf {\bibinfo {volume} {108}},\ \bibinfo {pages}
  {124062} (\bibinfo {year} {2023})},\ \Eprint
  {http://arxiv.org/abs/2309.06498} {arXiv:2309.06498 [gr-qc]} \BibitemShut
  {NoStop}%
\bibitem [{\citenamefont {Speeney}\ \emph {et~al.}(2022)\citenamefont
  {Speeney}, \citenamefont {Antonelli}, \citenamefont {Baibhav},\ and\
  \citenamefont {Berti}}]{Speeney:2022ryg}%
  \BibitemOpen
  \bibfield  {author} {\bibinfo {author} {\bibfnamefont {Nicholas}\
  \bibnamefont {Speeney}}, \bibinfo {author} {\bibfnamefont {Andrea}\
  \bibnamefont {Antonelli}}, \bibinfo {author} {\bibfnamefont {Vishal}\
  \bibnamefont {Baibhav}}, \ and\ \bibinfo {author} {\bibfnamefont {Emanuele}\
  \bibnamefont {Berti}},\ }\bibfield  {title} {\enquote {\bibinfo {title}
  {{Impact of relativistic corrections on the detectability of dark-matter
  spikes with gravitational waves}},}\ }\href {\doibase
  10.1103/PhysRevD.106.044027} {\bibfield  {journal} {\bibinfo  {journal}
  {Phys. Rev. D}\ }\textbf {\bibinfo {volume} {106}},\ \bibinfo {pages}
  {044027} (\bibinfo {year} {2022})},\ \Eprint
  {http://arxiv.org/abs/2204.12508} {arXiv:2204.12508 [gr-qc]} \BibitemShut
  {NoStop}%
\bibitem [{\citenamefont {Montalvo}\ \emph {et~al.}(2024)\citenamefont
  {Montalvo}, \citenamefont {Smith-Orlik}, \citenamefont {Rastgoo},
  \citenamefont {Sagunski}, \citenamefont {Becker},\ and\ \citenamefont
  {Khan}}]{Montalvo:2024iwq}%
  \BibitemOpen
  \bibfield  {author} {\bibinfo {author} {\bibfnamefont {Diego}\ \bibnamefont
  {Montalvo}}, \bibinfo {author} {\bibfnamefont {Adam}\ \bibnamefont
  {Smith-Orlik}}, \bibinfo {author} {\bibfnamefont {Saeed}\ \bibnamefont
  {Rastgoo}}, \bibinfo {author} {\bibfnamefont {Laura}\ \bibnamefont
  {Sagunski}}, \bibinfo {author} {\bibfnamefont {Niklas}\ \bibnamefont
  {Becker}}, \ and\ \bibinfo {author} {\bibfnamefont {Hazkeel}\ \bibnamefont
  {Khan}},\ }\bibfield  {title} {\enquote {\bibinfo {title} {{Post-Newtonian
  effects in compact binaries with a dark matter spike: A Lagrangian
  approach}},}\ }\href@noop {} {\  (\bibinfo {year} {2024})},\ \Eprint
  {http://arxiv.org/abs/2401.06084} {arXiv:2401.06084 [gr-qc]} \BibitemShut
  {NoStop}%
\bibitem [{\citenamefont {Yue}\ and\ \citenamefont {Cao}(2019)}]{Yue:2019ozq}%
  \BibitemOpen
  \bibfield  {author} {\bibinfo {author} {\bibfnamefont {Xiao-Jun}\
  \bibnamefont {Yue}}\ and\ \bibinfo {author} {\bibfnamefont {Zhoujian}\
  \bibnamefont {Cao}},\ }\bibfield  {title} {\enquote {\bibinfo {title} {{Dark
  matter minispike: A significant enhancement of eccentricity for
  intermediate-mass-ratio inspirals}},}\ }\href {\doibase
  10.1103/PhysRevD.100.043013} {\bibfield  {journal} {\bibinfo  {journal}
  {Phys. Rev. D}\ }\textbf {\bibinfo {volume} {100}},\ \bibinfo {pages}
  {043013} (\bibinfo {year} {2019})},\ \Eprint
  {http://arxiv.org/abs/1908.10241} {arXiv:1908.10241 [astro-ph.HE]}
  \BibitemShut {NoStop}%
\bibitem [{\citenamefont {Tang}\ and\ \citenamefont
  {Wang}(2021)}]{Tang:2020jhx}%
  \BibitemOpen
  \bibfield  {author} {\bibinfo {author} {\bibfnamefont {Meirong}\ \bibnamefont
  {Tang}}\ and\ \bibinfo {author} {\bibfnamefont {Jiancheng}\ \bibnamefont
  {Wang}},\ }\bibfield  {title} {\enquote {\bibinfo {title} {{The eccentricity
  enhancement effect of intermediate-mass-ratio-inspirals: dark matter and
  black hole mass}},}\ }\href {\doibase 10.1088/1674-1137/abc680} {\bibfield
  {journal} {\bibinfo  {journal} {Chin. Phys. C}\ }\textbf {\bibinfo {volume}
  {45}},\ \bibinfo {pages} {015110} (\bibinfo {year} {2021})},\ \Eprint
  {http://arxiv.org/abs/2005.11933} {arXiv:2005.11933 [gr-qc]} \BibitemShut
  {NoStop}%
\bibitem [{\citenamefont {Li}\ \emph {et~al.}(2022)\citenamefont {Li},
  \citenamefont {Tang},\ and\ \citenamefont {Wu}}]{Li2}%
  \BibitemOpen
  \bibfield  {author} {\bibinfo {author} {\bibfnamefont {Gen-Liang}\
  \bibnamefont {Li}}, \bibinfo {author} {\bibfnamefont {Yong}\ \bibnamefont
  {Tang}}, \ and\ \bibinfo {author} {\bibfnamefont {Yue-Liang}\ \bibnamefont
  {Wu}},\ }\bibfield  {title} {\enquote {\bibinfo {title} {{Probing dark matter
  spikes via gravitational waves of extreme-mass-ratio inspirals}},}\ }\href
  {\doibase 10.1007/s11433-022-1930-9} {\bibfield  {journal} {\bibinfo
  {journal} {Sci. China Phys. Mech. Astron.}\ }\textbf {\bibinfo {volume}
  {65}},\ \bibinfo {pages} {100412} (\bibinfo {year} {2022})},\ \Eprint
  {http://arxiv.org/abs/2112.14041} {arXiv:2112.14041 [astro-ph.CO]}
  \BibitemShut {NoStop}%
\bibitem [{\citenamefont {Becker}\ \emph {et~al.}(2022)\citenamefont {Becker},
  \citenamefont {Sagunski}, \citenamefont {Prinz},\ and\ \citenamefont
  {Rastgoo}}]{Becker:2021ivq}%
  \BibitemOpen
  \bibfield  {author} {\bibinfo {author} {\bibfnamefont {Niklas}\ \bibnamefont
  {Becker}}, \bibinfo {author} {\bibfnamefont {Laura}\ \bibnamefont
  {Sagunski}}, \bibinfo {author} {\bibfnamefont {Lukas}\ \bibnamefont {Prinz}},
  \ and\ \bibinfo {author} {\bibfnamefont {Saeed}\ \bibnamefont {Rastgoo}},\
  }\bibfield  {title} {\enquote {\bibinfo {title} {{Circularization versus
  eccentrification in intermediate mass ratio inspirals inside dark matter
  spikes}},}\ }\href {\doibase 10.1103/PhysRevD.105.063029} {\bibfield
  {journal} {\bibinfo  {journal} {Phys. Rev. D}\ }\textbf {\bibinfo {volume}
  {105}},\ \bibinfo {pages} {063029} (\bibinfo {year} {2022})},\ \Eprint
  {http://arxiv.org/abs/2112.09586} {arXiv:2112.09586 [gr-qc]} \BibitemShut
  {NoStop}%
\bibitem [{\citenamefont {{Hu}}\ \emph {et~al.}(2023)\citenamefont {{Hu}},
  \citenamefont {{Cai}},\ and\ \citenamefont {{Wang}}}]{2023arXiv231214041H}%
  \BibitemOpen
  \bibfield  {author} {\bibinfo {author} {\bibfnamefont {Li}~\bibnamefont
  {{Hu}}}, \bibinfo {author} {\bibfnamefont {Rong-Gen}\ \bibnamefont {{Cai}}},
  \ and\ \bibinfo {author} {\bibfnamefont {Shao-Jiang}\ \bibnamefont
  {{Wang}}},\ }\bibfield  {title} {\enquote {\bibinfo {title} {{Distinctive
  GWBs from eccentric inspiraling SMBH binaries with a DM spike}},}\ }\href
  {\doibase 10.48550/arXiv.2312.14041} {\bibfield  {journal} {\bibinfo
  {journal} {arXiv e-prints}\ ,\ \bibinfo {eid} {arXiv:2312.14041}} (\bibinfo
  {year} {2023})},\ \Eprint {http://arxiv.org/abs/2312.14041} {arXiv:2312.14041
  [gr-qc]} \BibitemShut {NoStop}%
\bibitem [{\citenamefont {Mukherjee}\ \emph {et~al.}(2023)\citenamefont
  {Mukherjee}, \citenamefont {Holgado}, \citenamefont {Ogiya},\ and\
  \citenamefont {Trac}}]{Mukherjee:2023lzn}%
  \BibitemOpen
  \bibfield  {author} {\bibinfo {author} {\bibfnamefont {Diptajyoti}\
  \bibnamefont {Mukherjee}}, \bibinfo {author} {\bibfnamefont {A.~Miguel}\
  \bibnamefont {Holgado}}, \bibinfo {author} {\bibfnamefont {Go}~\bibnamefont
  {Ogiya}}, \ and\ \bibinfo {author} {\bibfnamefont {Hy}~\bibnamefont {Trac}},\
  }\bibfield  {title} {\enquote {\bibinfo {title} {{Examining the Effects of
  Dark Matter Spikes on Eccentric Intermediate Mass Ratio Inspirals Using
  $N$-body Simulations}},}\ }\href@noop {} {\  (\bibinfo {year} {2023})},\
  \Eprint {http://arxiv.org/abs/2312.02275} {arXiv:2312.02275 [astro-ph.CO]}
  \BibitemShut {NoStop}%
\bibitem [{\citenamefont {Chowdhuri}\ \emph {et~al.}(2023)\citenamefont
  {Chowdhuri}, \citenamefont {Singh}, \citenamefont {Kangsabanik},\ and\
  \citenamefont {Bhattacharyya}}]{AbhishekChowdhuri:2023cle}%
  \BibitemOpen
  \bibfield  {author} {\bibinfo {author} {\bibfnamefont {Abhishek}\
  \bibnamefont {Chowdhuri}}, \bibinfo {author} {\bibfnamefont {Rishabh~Kumar}\
  \bibnamefont {Singh}}, \bibinfo {author} {\bibfnamefont {Kaushik}\
  \bibnamefont {Kangsabanik}}, \ and\ \bibinfo {author} {\bibfnamefont {Arpan}\
  \bibnamefont {Bhattacharyya}},\ }\bibfield  {title} {\enquote {\bibinfo
  {title} {{Gravitational radiation from hyperbolic encounters in the presence
  of dark matter}},}\ }\href@noop {} {\  (\bibinfo {year} {2023})},\ \Eprint
  {http://arxiv.org/abs/2306.11787} {arXiv:2306.11787 [gr-qc]} \BibitemShut
  {NoStop}%
\bibitem [{\citenamefont
  {Chandrasekhar}(1943{\natexlab{a}})}]{Chandrasekhar1943a}%
  \BibitemOpen
  \bibfield  {author} {\bibinfo {author} {\bibfnamefont {S.}~\bibnamefont
  {Chandrasekhar}},\ }\bibfield  {title} {\enquote {\bibinfo {title} {Dynamical
  friction. {I}. general considerations: the coefficient of dynamical
  friction.}}\ }\href {\doibase 10.1086/144517} {\bibfield  {journal} {\bibinfo
   {journal} {The Astrophysical Journal}\ }\textbf {\bibinfo {volume} {97}},\
  \bibinfo {pages} {255} (\bibinfo {year} {1943}{\natexlab{a}})}\BibitemShut
  {NoStop}%
\bibitem [{\citenamefont
  {Chandrasekhar}(1943{\natexlab{b}})}]{Chandrasekhar1943b}%
  \BibitemOpen
  \bibfield  {author} {\bibinfo {author} {\bibfnamefont {S.}~\bibnamefont
  {Chandrasekhar}},\ }\bibfield  {title} {\enquote {\bibinfo {title} {Dynamical
  friction. {II}. the rate of escape of stars from clusters and the evidence
  for the operation of dynamical friction.}}\ }\href {\doibase 10.1086/144518}
  {\bibfield  {journal} {\bibinfo  {journal} {The Astrophysical Journal}\
  }\textbf {\bibinfo {volume} {97}},\ \bibinfo {pages} {263} (\bibinfo {year}
  {1943}{\natexlab{b}})}\BibitemShut {NoStop}%
\bibitem [{\citenamefont
  {Chandrasekhar}(1943{\natexlab{c}})}]{Chandrasekhar1943c}%
  \BibitemOpen
  \bibfield  {author} {\bibinfo {author} {\bibfnamefont {S.}~\bibnamefont
  {Chandrasekhar}},\ }\bibfield  {title} {\enquote {\bibinfo {title} {Dynamical
  friction. {III}. a more exact theory of the rate of escape of stars from
  clusters.}}\ }\href {\doibase 10.1086/144544} {\bibfield  {journal} {\bibinfo
   {journal} {The Astrophysical Journal}\ }\textbf {\bibinfo {volume} {98}},\
  \bibinfo {pages} {54} (\bibinfo {year} {1943}{\natexlab{c}})}\BibitemShut
  {NoStop}%
\bibitem [{\citenamefont {Dosopoulou}(2023)}]{Dosopoulou:2023umg}%
  \BibitemOpen
  \bibfield  {author} {\bibinfo {author} {\bibfnamefont {Fani}\ \bibnamefont
  {Dosopoulou}},\ }\bibfield  {title} {\enquote {\bibinfo {title} {{Dynamical
  friction in dark matter spikes: corrections to Chandrasekhar's formula}},}\
  }\href@noop {} {\  (\bibinfo {year} {2023})},\ \Eprint
  {http://arxiv.org/abs/2305.17281} {arXiv:2305.17281 [astro-ph.HE]}
  \BibitemShut {NoStop}%
\bibitem [{\citenamefont {Kavanagh}\ \emph {et~al.}(2020)\citenamefont
  {Kavanagh}, \citenamefont {Nichols}, \citenamefont {Bertone},\ and\
  \citenamefont {Gaggero}}]{Kavanagh:2020cfn}%
  \BibitemOpen
  \bibfield  {author} {\bibinfo {author} {\bibfnamefont {Bradley~J.}\
  \bibnamefont {Kavanagh}}, \bibinfo {author} {\bibfnamefont {David~A.}\
  \bibnamefont {Nichols}}, \bibinfo {author} {\bibfnamefont {Gianfranco}\
  \bibnamefont {Bertone}}, \ and\ \bibinfo {author} {\bibfnamefont {Daniele}\
  \bibnamefont {Gaggero}},\ }\bibfield  {title} {\enquote {\bibinfo {title}
  {{Detecting dark matter around black holes with gravitational waves: Effects
  of dark-matter dynamics on the gravitational waveform}},}\ }\href {\doibase
  10.1103/PhysRevD.102.083006} {\bibfield  {journal} {\bibinfo  {journal}
  {Phys. Rev. D}\ }\textbf {\bibinfo {volume} {102}},\ \bibinfo {pages}
  {083006} (\bibinfo {year} {2020})},\ \Eprint
  {http://arxiv.org/abs/2002.12811} {arXiv:2002.12811 [gr-qc]} \BibitemShut
  {NoStop}%
\bibitem [{\citenamefont {Kavanagh}(2024)}]{NbodyIMRI}%
  \BibitemOpen
  \bibfield  {author} {\bibinfo {author} {\bibfnamefont {Bradley~J.}\
  \bibnamefont {Kavanagh}},\ }\href {\doibase 10.5281/zenodo.10641173}
  {\enquote {\bibinfo {title} {{NbodyIMRI [Code, v1.0]}},}\ }\bibinfo
  {howpublished} {\url{https://github.com/bradkav/NbodyIMRI},
  \href{https://doi.org/10.5281/zenodo.10641173}{DOI:10.5281/zenodo.10641173}}
  (\bibinfo {year} {2024})\BibitemShut {NoStop}%
\bibitem [{\citenamefont {Kavanagh}(2020)}]{HaloFeedback}%
  \BibitemOpen
  \bibfield  {author} {\bibinfo {author} {\bibfnamefont {Bradley~J.}\
  \bibnamefont {Kavanagh}},\ }\href {\doibase 10.5281/zenodo.3688813} {\enquote
  {\bibinfo {title} {{HaloFeedback [Code, v0.9]}},}\ }\bibinfo {howpublished}
  {\url{https://github.com/bradkav/HaloFeedback},
  \href{https://doi.org/10.5281/zenodo.3688813}{DOI:10.5281/zenodo.3688813}}
  (\bibinfo {year} {2020})\BibitemShut {NoStop}%
\bibitem [{\citenamefont {Karydas}\ \emph {et~al.}(2024)\citenamefont
  {Karydas}, \citenamefont {Kavanagh},\ and\ \citenamefont
  {Bertone}}]{BetterSpikesI}%
  \BibitemOpen
  \bibfield  {author} {\bibinfo {author} {\bibfnamefont {Theophanes~K.}\
  \bibnamefont {Karydas}}, \bibinfo {author} {\bibfnamefont {Bradley~J.}\
  \bibnamefont {Kavanagh}}, \ and\ \bibinfo {author} {\bibfnamefont
  {Gianfranco}\ \bibnamefont {Bertone}},\ }\bibfield  {title} {\enquote
  {\bibinfo {title} {{Sharpening the dark matter signature in gravitational
  waveforms I: Accretion and eccentricity evolution}},}\ }\href@noop {} {\
  (\bibinfo {year} {2024})},\ \Eprint {http://arxiv.org/abs/2402.13053}
  {arXiv:2402.13053 [gr-qc]} \BibitemShut {NoStop}%
\bibitem [{\citenamefont {{van Albada}}\ and\ \citenamefont
  {{Szomoru}}(2020)}]{2020IAUS..351..532V}%
  \BibitemOpen
  \bibfield  {author} {\bibinfo {author} {\bibfnamefont {Tjeerd~S.}\
  \bibnamefont {{van Albada}}}\ and\ \bibinfo {author} {\bibfnamefont {Arpad}\
  \bibnamefont {{Szomoru}}},\ }\bibfield  {title} {\enquote {\bibinfo {title}
  {{The Chandrasekhar Spitzer controversy and the (ir)relevance of distant
  interactions}},}\ }in\ \href {\doibase 10.1017/S1743921319006768} {\emph
  {\bibinfo {booktitle} {Star Clusters: From the Milky Way to the Early
  Universe}}},\ Vol.\ \bibinfo {volume} {351},\ \bibinfo {editor} {edited by\
  \bibinfo {editor} {\bibfnamefont {Angela}\ \bibnamefont {{Bragaglia}}},
  \bibinfo {editor} {\bibfnamefont {Melvyn}\ \bibnamefont {{Davies}}}, \bibinfo
  {editor} {\bibfnamefont {Alison}\ \bibnamefont {{Sills}}}, \ and\ \bibinfo
  {editor} {\bibfnamefont {Enrico}\ \bibnamefont {{Vesperini}}}}\ (\bibinfo
  {year} {2020})\ pp.\ \bibinfo {pages} {532--535}\BibitemShut {NoStop}%
\bibitem [{\citenamefont {{Mulder}}(1983)}]{1983A&A...117....9M}%
  \BibitemOpen
  \bibfield  {author} {\bibinfo {author} {\bibfnamefont {W.~A.}\ \bibnamefont
  {{Mulder}}},\ }\bibfield  {title} {\enquote {\bibinfo {title} {{Dynamical
  friction on extended objects}},}\ }\href@noop {} {\bibfield  {journal}
  {\bibinfo  {journal} {A\& A}\ }\textbf {\bibinfo {volume} {117}},\ \bibinfo
  {pages} {9--16} (\bibinfo {year} {1983})}\BibitemShut {NoStop}%
\bibitem [{\citenamefont {{Ostriker}}(1999)}]{1999ApJ...513..252O}%
  \BibitemOpen
  \bibfield  {author} {\bibinfo {author} {\bibfnamefont {Eve~C.}\ \bibnamefont
  {{Ostriker}}},\ }\bibfield  {title} {\enquote {\bibinfo {title} {{Dynamical
  Friction in a Gaseous Medium}},}\ }\href {\doibase 10.1086/306858} {\bibfield
   {journal} {\bibinfo  {journal} {ApJ}\ }\textbf {\bibinfo {volume} {513}},\
  \bibinfo {pages} {252--258} (\bibinfo {year} {1999})},\ \Eprint
  {http://arxiv.org/abs/astro-ph/9810324} {arXiv:astro-ph/9810324 [astro-ph]}
  \BibitemShut {NoStop}%
\bibitem [{\citenamefont {Athanassoula}\ \emph {et~al.}(2000)\citenamefont
  {Athanassoula}, \citenamefont {Fady}, \citenamefont {Lambert},\ and\
  \citenamefont {Bosma}}]{Athanassoula:1999wz}%
  \BibitemOpen
  \bibfield  {author} {\bibinfo {author} {\bibfnamefont {E.}~\bibnamefont
  {Athanassoula}}, \bibinfo {author} {\bibfnamefont {E.}~\bibnamefont {Fady}},
  \bibinfo {author} {\bibfnamefont {J.~C.}\ \bibnamefont {Lambert}}, \ and\
  \bibinfo {author} {\bibfnamefont {A.}~\bibnamefont {Bosma}},\ }\bibfield
  {title} {\enquote {\bibinfo {title} {{Optimal softening for force
  calculations in collisionless n-body simulations}},}\ }\href {\doibase
  10.1046/j.1365-8711.2000.03316.x} {\bibfield  {journal} {\bibinfo  {journal}
  {Mon. Not. Roy. Astron. Soc.}\ }\textbf {\bibinfo {volume} {314}},\ \bibinfo
  {pages} {475} (\bibinfo {year} {2000})},\ \Eprint
  {http://arxiv.org/abs/astro-ph/9912467} {arXiv:astro-ph/9912467} \BibitemShut
  {NoStop}%
\bibitem [{\citenamefont {Dehnen}(2001)}]{Dehnen:2000nh}%
  \BibitemOpen
  \bibfield  {author} {\bibinfo {author} {\bibfnamefont {Walter}\ \bibnamefont
  {Dehnen}},\ }\bibfield  {title} {\enquote {\bibinfo {title} {{Towards optimal
  softening in 3-D n-body codes: I. minimizing the force error}},}\ }\href
  {\doibase 10.1046/j.1365-8711.2001.04237.x} {\bibfield  {journal} {\bibinfo
  {journal} {Mon. Not. Roy. Astron. Soc.}\ }\textbf {\bibinfo {volume} {324}},\
  \bibinfo {pages} {273} (\bibinfo {year} {2001})},\ \Eprint
  {http://arxiv.org/abs/astro-ph/0011568} {arXiv:astro-ph/0011568} \BibitemShut
  {NoStop}%
\bibitem [{\citenamefont {{Pfenniger}}\ and\ \citenamefont
  {{Friedli}}(1993)}]{1993A&A...270..561P}%
  \BibitemOpen
  \bibfield  {author} {\bibinfo {author} {\bibfnamefont {D.}~\bibnamefont
  {{Pfenniger}}}\ and\ \bibinfo {author} {\bibfnamefont {D.}~\bibnamefont
  {{Friedli}}},\ }\bibfield  {title} {\enquote {\bibinfo {title}
  {{Computational issues connected with 3D N-body simulations}},}\ }\href@noop
  {} {\bibfield  {journal} {\bibinfo  {journal} {Astronomy \& Astrophysics}\
  }\textbf {\bibinfo {volume} {270}},\ \bibinfo {pages} {561--572} (\bibinfo
  {year} {1993})}\BibitemShut {NoStop}%
\bibitem [{\citenamefont {{Kinoshita}}\ \emph {et~al.}(1990)\citenamefont
  {{Kinoshita}}, \citenamefont {{Yoshida}},\ and\ \citenamefont
  {{Nakai}}}]{1990CeMDA..50...59K}%
  \BibitemOpen
  \bibfield  {author} {\bibinfo {author} {\bibfnamefont {Hiroshi}\ \bibnamefont
  {{Kinoshita}}}, \bibinfo {author} {\bibfnamefont {Haruo}\ \bibnamefont
  {{Yoshida}}}, \ and\ \bibinfo {author} {\bibfnamefont {Hiroshi}\ \bibnamefont
  {{Nakai}}},\ }\bibfield  {title} {\enquote {\bibinfo {title} {{Symplectic
  integrators and their application to dynamical astronomy}},}\ }\href
  {\doibase 10.1007/BF00048986} {\bibfield  {journal} {\bibinfo  {journal}
  {Celestial Mechanics and Dynamical Astronomy}\ }\textbf {\bibinfo {volume}
  {50}},\ \bibinfo {pages} {59--71} (\bibinfo {year} {1990})}\BibitemShut
  {NoStop}%
\bibitem [{\citenamefont {{Mei}}\ and\ \citenamefont
  {{Wu}}(2017)}]{2017JCoPh.338..567M}%
  \BibitemOpen
  \bibfield  {author} {\bibinfo {author} {\bibfnamefont {Lijie}\ \bibnamefont
  {{Mei}}}\ and\ \bibinfo {author} {\bibfnamefont {Xinyuan}\ \bibnamefont
  {{Wu}}},\ }\bibfield  {title} {\enquote {\bibinfo {title} {{Symplectic
  exponential Runge-Kutta methods for solving nonlinear Hamiltonian
  systems}},}\ }\href {\doibase 10.1016/j.jcp.2017.03.018} {\bibfield
  {journal} {\bibinfo  {journal} {Journal of Computational Physics}\ }\textbf
  {\bibinfo {volume} {338}},\ \bibinfo {pages} {567--584} (\bibinfo {year}
  {2017})}\BibitemShut {NoStop}%
\bibitem [{\citenamefont {Verlet}(1967)}]{Verlet1967}%
  \BibitemOpen
  \bibfield  {author} {\bibinfo {author} {\bibfnamefont {Loup}\ \bibnamefont
  {Verlet}},\ }\bibfield  {title} {\enquote {\bibinfo {title} {Computer
  "experiments" on classical fluids. i. thermodynamical properties of
  lennard-jones molecules},}\ }\href {\doibase 10.1103/physrev.159.98}
  {\bibfield  {journal} {\bibinfo  {journal} {Physical Review}\ }\textbf
  {\bibinfo {volume} {159}},\ \bibinfo {pages} {98--103} (\bibinfo {year}
  {1967})}\BibitemShut {NoStop}%
\bibitem [{\citenamefont {{Forest}}\ and\ \citenamefont
  {{Ruth}}(1990)}]{1990PhyD...43..105F}%
  \BibitemOpen
  \bibfield  {author} {\bibinfo {author} {\bibfnamefont {Etienne}\ \bibnamefont
  {{Forest}}}\ and\ \bibinfo {author} {\bibfnamefont {Ronald~D.}\ \bibnamefont
  {{Ruth}}},\ }\bibfield  {title} {\enquote {\bibinfo {title} {{Fourth-order
  symplectic integration}},}\ }\href {\doibase 10.1016/0167-2789(90)90019-L}
  {\bibfield  {journal} {\bibinfo  {journal} {Physica D Nonlinear Phenomena}\
  }\textbf {\bibinfo {volume} {43}},\ \bibinfo {pages} {105--117} (\bibinfo
  {year} {1990})}\BibitemShut {NoStop}%
\bibitem [{\citenamefont {{Yoshida}}(1990)}]{1990PhLA..150..262Y}%
  \BibitemOpen
  \bibfield  {author} {\bibinfo {author} {\bibfnamefont {Haruo}\ \bibnamefont
  {{Yoshida}}},\ }\bibfield  {title} {\enquote {\bibinfo {title} {{Construction
  of higher order symplectic integrators}},}\ }\href {\doibase
  10.1016/0375-9601(90)90092-3} {\bibfield  {journal} {\bibinfo  {journal}
  {Physics Letters A}\ }\textbf {\bibinfo {volume} {150}},\ \bibinfo {pages}
  {262--268} (\bibinfo {year} {1990})}\BibitemShut {NoStop}%
\bibitem [{\citenamefont {{Omelyan}}\ \emph {et~al.}(2002)\citenamefont
  {{Omelyan}}, \citenamefont {{Mryglod}},\ and\ \citenamefont
  {{Folk}}}]{2002CoPhC.146..188O}%
  \BibitemOpen
  \bibfield  {author} {\bibinfo {author} {\bibfnamefont {I.~P.}\ \bibnamefont
  {{Omelyan}}}, \bibinfo {author} {\bibfnamefont {I.~M.}\ \bibnamefont
  {{Mryglod}}}, \ and\ \bibinfo {author} {\bibfnamefont {R.}~\bibnamefont
  {{Folk}}},\ }\bibfield  {title} {\enquote {\bibinfo {title} {{Optimized
  Forest-Ruth- and Suzuki-like algorithms for integration of motion in
  many-body systems}},}\ }\href {\doibase 10.1016/S0010-4655(02)00451-4}
  {\bibfield  {journal} {\bibinfo  {journal} {Computer Physics Communications}\
  }\textbf {\bibinfo {volume} {146}},\ \bibinfo {pages} {188--202} (\bibinfo
  {year} {2002})},\ \Eprint {http://arxiv.org/abs/cond-mat/0110585}
  {arXiv:cond-mat/0110585 [cond-mat.stat-mech]} \BibitemShut {NoStop}%
\bibitem [{\citenamefont {{Baker}}\ and\ \citenamefont
  {{Centrella}}(2005)}]{2005CQGra..22S.355B}%
  \BibitemOpen
  \bibfield  {author} {\bibinfo {author} {\bibfnamefont {J.}~\bibnamefont
  {{Baker}}}\ and\ \bibinfo {author} {\bibfnamefont {J.}~\bibnamefont
  {{Centrella}}},\ }\bibfield  {title} {\enquote {\bibinfo {title} {{Impact of
  LISA's low-frequency sensitivity on observations of massive black-hole
  mergers}},}\ }\href {\doibase 10.1088/0264-9381/22/10/029} {\bibfield
  {journal} {\bibinfo  {journal} {Classical and Quantum Gravity}\ }\textbf
  {\bibinfo {volume} {22}},\ \bibinfo {pages} {S355--S362} (\bibinfo {year}
  {2005})},\ \Eprint {http://arxiv.org/abs/astro-ph/0411616}
  {arXiv:astro-ph/0411616 [astro-ph]} \BibitemShut {NoStop}%
\bibitem [{\citenamefont {{Binney}}\ and\ \citenamefont
  {{Tremaine}}(2008)}]{BinneyAndTremaine}%
  \BibitemOpen
  \bibfield  {author} {\bibinfo {author} {\bibfnamefont {J.}~\bibnamefont
  {{Binney}}}\ and\ \bibinfo {author} {\bibfnamefont {S.}~\bibnamefont
  {{Tremaine}}},\ }\href@noop {} {\emph {\bibinfo {title} {{Galactic Dynamics:
  Second Edition}}}}\ (\bibinfo  {publisher} {Princeton University Press},\
  \bibinfo {year} {2008})\BibitemShut {NoStop}%
\bibitem [{\citenamefont {Lacroix}\ \emph {et~al.}(2018)\citenamefont
  {Lacroix}, \citenamefont {Stref},\ and\ \citenamefont
  {Lavalle}}]{Lacroix:2018qqh}%
  \BibitemOpen
  \bibfield  {author} {\bibinfo {author} {\bibfnamefont {Thomas}\ \bibnamefont
  {Lacroix}}, \bibinfo {author} {\bibfnamefont {Martin}\ \bibnamefont {Stref}},
  \ and\ \bibinfo {author} {\bibfnamefont {Julien}\ \bibnamefont {Lavalle}},\
  }\bibfield  {title} {\enquote {\bibinfo {title} {{Anatomy of Eddington-like
  inversion methods in the context of dark matter searches}},}\ }\href
  {\doibase 10.1088/1475-7516/2018/09/040} {\bibfield  {journal} {\bibinfo
  {journal} {JCAP}\ }\textbf {\bibinfo {volume} {09}},\ \bibinfo {pages} {040}
  (\bibinfo {year} {2018})},\ \Eprint {http://arxiv.org/abs/1805.02403}
  {arXiv:1805.02403 [astro-ph.GA]} \BibitemShut {NoStop}%
\bibitem [{\citenamefont {{Osipkov}}(1979)}]{1979SvAL....5...42O}%
  \BibitemOpen
  \bibfield  {author} {\bibinfo {author} {\bibfnamefont {L.~P.}\ \bibnamefont
  {{Osipkov}}},\ }\bibfield  {title} {\enquote {\bibinfo {title} {{Spherical
  systems of gravitating bodies with an ellipsoidal velocity distribution}},}\
  }\href@noop {} {\bibfield  {journal} {\bibinfo  {journal} {Soviet Astronomy
  Letters}\ }\textbf {\bibinfo {volume} {5}},\ \bibinfo {pages} {42--44}
  (\bibinfo {year} {1979})}\BibitemShut {NoStop}%
\bibitem [{\citenamefont {{Merritt}}(1985)}]{1985AJ.....90.1027M}%
  \BibitemOpen
  \bibfield  {author} {\bibinfo {author} {\bibfnamefont {D.}~\bibnamefont
  {{Merritt}}},\ }\bibfield  {title} {\enquote {\bibinfo {title} {{Spherical
  stellar systems with spheroidal velocity distributions}},}\ }\href {\doibase
  10.1086/113810} {\bibfield  {journal} {\bibinfo  {journal} {Astron. Jour.}\
  }\textbf {\bibinfo {volume} {90}},\ \bibinfo {pages} {1027--1037} (\bibinfo
  {year} {1985})}\BibitemShut {NoStop}%
\bibitem [{\citenamefont {{Binney}}(1977)}]{1977MNRAS.181..735B}%
  \BibitemOpen
  \bibfield  {author} {\bibinfo {author} {\bibfnamefont {J.}~\bibnamefont
  {{Binney}}},\ }\bibfield  {title} {\enquote {\bibinfo {title} {{Dynamical
  friction in aspherical clusters.}}}\ }\href {\doibase
  10.1093/mnras/181.4.735} {\bibfield  {journal} {\bibinfo  {journal} {Mon.
  Not. Roy. Astron. Soc.}\ }\textbf {\bibinfo {volume} {181}},\ \bibinfo
  {pages} {735--746} (\bibinfo {year} {1977})}\BibitemShut {NoStop}%
\bibitem [{\citenamefont {{Syer}}(1994)}]{1994MNRAS.270..205S}%
  \BibitemOpen
  \bibfield  {author} {\bibinfo {author} {\bibfnamefont {D.}~\bibnamefont
  {{Syer}}},\ }\bibfield  {title} {\enquote {\bibinfo {title} {{Relativistic
  Dynamical Friction in the Weak Scattering Limit}},}\ }\href {\doibase
  10.1093/mnras/270.1.205} {\bibfield  {journal} {\bibinfo  {journal} {mnras}\
  }\textbf {\bibinfo {volume} {270}},\ \bibinfo {pages} {205} (\bibinfo {year}
  {1994})}\BibitemShut {NoStop}%
\bibitem [{\citenamefont {{Chiari}}\ and\ \citenamefont {{Di
  Cintio}}(2023)}]{2023A&A...677A.140C}%
  \BibitemOpen
  \bibfield  {author} {\bibinfo {author} {\bibfnamefont {Caterina}\
  \bibnamefont {{Chiari}}}\ and\ \bibinfo {author} {\bibfnamefont
  {Pierfrancesco}\ \bibnamefont {{Di Cintio}}},\ }\bibfield  {title} {\enquote
  {\bibinfo {title} {{Relativistic dynamical friction in stellar systems}},}\
  }\href {\doibase 10.1051/0004-6361/202245569} {\bibfield  {journal} {\bibinfo
   {journal} {Astronomy \& Astrophysics}\ }\textbf {\bibinfo {volume} {677}},\
  \bibinfo {eid} {A140} (\bibinfo {year} {2023})},\ \Eprint
  {http://arxiv.org/abs/2207.05728} {arXiv:2207.05728 [astro-ph.GA]}
  \BibitemShut {NoStop}%
\bibitem [{\citenamefont {{Chiron}}\ and\ \citenamefont
  {{Marcos}}(2019)}]{2019JMP....60e2901C}%
  \BibitemOpen
  \bibfield  {author} {\bibinfo {author} {\bibfnamefont {D.}~\bibnamefont
  {{Chiron}}}\ and\ \bibinfo {author} {\bibfnamefont {B.}~\bibnamefont
  {{Marcos}}},\ }\bibfield  {title} {\enquote {\bibinfo {title} {{Series
  expansions of the deflection angle in the scattering problem for power-law
  potentials}},}\ }\href {\doibase 10.1063/1.5055713} {\bibfield  {journal}
  {\bibinfo  {journal} {Journal of Mathematical Physics}\ }\textbf {\bibinfo
  {volume} {60}},\ \bibinfo {eid} {052901} (\bibinfo {year}
  {2019})}\BibitemShut {NoStop}%
\bibitem [{\citenamefont {Maggiore}(2007)}]{Maggiore:2007ulw}%
  \BibitemOpen
  \bibfield  {author} {\bibinfo {author} {\bibfnamefont {Michele}\ \bibnamefont
  {Maggiore}},\ }\href@noop {} {\emph {\bibinfo {title} {{Gravitational Waves.
  Vol. 1: Theory and Experiments}}}},\ Oxford Master Series in Physics\
  (\bibinfo  {publisher} {Oxford University Press},\ \bibinfo {year}
  {2007})\BibitemShut {NoStop}%
\bibitem [{\citenamefont {{Dermott}}\ and\ \citenamefont
  {{Murray}}(1983)}]{1983Natur.301..201D}%
  \BibitemOpen
  \bibfield  {author} {\bibinfo {author} {\bibfnamefont {S.~F.}\ \bibnamefont
  {{Dermott}}}\ and\ \bibinfo {author} {\bibfnamefont {C.~D.}\ \bibnamefont
  {{Murray}}},\ }\bibfield  {title} {\enquote {\bibinfo {title} {{Nature of the
  Kirkwood gaps in the asteroid belt}},}\ }\href {\doibase 10.1038/301201a0}
  {\bibfield  {journal} {\bibinfo  {journal} {\nat}\ }\textbf {\bibinfo
  {volume} {301}},\ \bibinfo {pages} {201--205} (\bibinfo {year}
  {1983})}\BibitemShut {NoStop}%
\bibitem [{\citenamefont {{Henrard}}\ and\ \citenamefont
  {{Lemaitre}}(1983)}]{1983CeMec..30..197H}%
  \BibitemOpen
  \bibfield  {author} {\bibinfo {author} {\bibfnamefont {J.}~\bibnamefont
  {{Henrard}}}\ and\ \bibinfo {author} {\bibfnamefont {A.}~\bibnamefont
  {{Lemaitre}}},\ }\bibfield  {title} {\enquote {\bibinfo {title} {{A Second
  Fundamental Model for Resonance}},}\ }\href {\doibase 10.1007/BF01234306}
  {\bibfield  {journal} {\bibinfo  {journal} {Celestial Mechanics}\ }\textbf
  {\bibinfo {volume} {30}},\ \bibinfo {pages} {197--218} (\bibinfo {year}
  {1983})}\BibitemShut {NoStop}%
\bibitem [{\citenamefont {{Winter}}\ and\ \citenamefont
  {{Murray}}(1997)}]{1997A&A...319..290W}%
  \BibitemOpen
  \bibfield  {author} {\bibinfo {author} {\bibfnamefont {O.~C.}\ \bibnamefont
  {{Winter}}}\ and\ \bibinfo {author} {\bibfnamefont {C.~D.}\ \bibnamefont
  {{Murray}}},\ }\bibfield  {title} {\enquote {\bibinfo {title} {{Resonance and
  chaos. I. First-order interior resonances.}}}\ }\href@noop {} {\bibfield
  {journal} {\bibinfo  {journal} {Astronomy \& Astrophysics}\ }\textbf
  {\bibinfo {volume} {319}},\ \bibinfo {pages} {290--304} (\bibinfo {year}
  {1997})}\BibitemShut {NoStop}%
\bibitem [{\citenamefont {{Hut}}\ \emph {et~al.}(1995)\citenamefont {{Hut}},
  \citenamefont {{Makino}},\ and\ \citenamefont
  {{McMillan}}}]{1995ApJ...443L..93H}%
  \BibitemOpen
  \bibfield  {author} {\bibinfo {author} {\bibfnamefont {Piet}\ \bibnamefont
  {{Hut}}}, \bibinfo {author} {\bibfnamefont {Jun}\ \bibnamefont {{Makino}}}, \
  and\ \bibinfo {author} {\bibfnamefont {Steve}\ \bibnamefont {{McMillan}}},\
  }\bibfield  {title} {\enquote {\bibinfo {title} {{Building a Better
  Leapfrog}},}\ }\href {\doibase 10.1086/187844} {\bibfield  {journal}
  {\bibinfo  {journal} {ApJ Letters}\ }\textbf {\bibinfo {volume} {443}},\
  \bibinfo {pages} {L93} (\bibinfo {year} {1995})}\BibitemShut {NoStop}%
\bibitem [{\citenamefont {{Di Cintio}}\ \emph {et~al.}(2017)\citenamefont {{Di
  Cintio}}, \citenamefont {{Ciotti}},\ and\ \citenamefont
  {{Nipoti}}}]{2017MNRAS.468.2222D}%
  \BibitemOpen
  \bibfield  {author} {\bibinfo {author} {\bibfnamefont {Pierfrancesco}\
  \bibnamefont {{Di Cintio}}}, \bibinfo {author} {\bibfnamefont {Luca}\
  \bibnamefont {{Ciotti}}}, \ and\ \bibinfo {author} {\bibfnamefont {Carlo}\
  \bibnamefont {{Nipoti}}},\ }\bibfield  {title} {\enquote {\bibinfo {title}
  {{Radially anisotropic systems with r$^{-{\ensuremath{\alpha}}}$ forces - II:
  radial-orbit instability}},}\ }\href {\doibase 10.1093/mnras/stx600}
  {\bibfield  {journal} {\bibinfo  {journal} {Mon. Not. Roy. Astron. Soc.}\
  }\textbf {\bibinfo {volume} {468}},\ \bibinfo {pages} {2222--2231} (\bibinfo
  {year} {2017})},\ \Eprint {http://arxiv.org/abs/1612.03603} {arXiv:1612.03603
  [astro-ph.GA]} \BibitemShut {NoStop}%
\bibitem [{\citenamefont {Lacki}\ and\ \citenamefont
  {Beacom}(2010)}]{Lacki:2010zf}%
  \BibitemOpen
  \bibfield  {author} {\bibinfo {author} {\bibfnamefont {Brian~C.}\
  \bibnamefont {Lacki}}\ and\ \bibinfo {author} {\bibfnamefont {John~F.}\
  \bibnamefont {Beacom}},\ }\bibfield  {title} {\enquote {\bibinfo {title}
  {{Primordial Black Holes as Dark Matter: Almost All or Almost Nothing}},}\
  }\href {\doibase 10.1088/2041-8205/720/1/L67} {\bibfield  {journal} {\bibinfo
   {journal} {Astrophys. J. Lett.}\ }\textbf {\bibinfo {volume} {720}},\
  \bibinfo {pages} {L67--L71} (\bibinfo {year} {2010})},\ \Eprint
  {http://arxiv.org/abs/1003.3466} {arXiv:1003.3466 [astro-ph.CO]} \BibitemShut
  {NoStop}%
\end{thebibliography}%
\end{document}